\documentclass[a4paper,11pt]{article}
\usepackage{pos}
\usepackage{mathdots}
\usepackage{tabularray}
\usepackage{tikz-cd}
\usetikzlibrary{calc} 
\usepackage{slashed}
\usepackage{nicefrac}
\usepackage{physics}
\usepackage{mathrsfs}
\usepackage{tikz}
\usepackage{hyperref}
\usepackage{url} 
\usepackage[dvipsnames]{xcolor}

\usepackage{genyoungtabtikz}
\Yboxdim{8pt}
\Ylinethick{1pt}
\newcommand{\md}{\mathrm{d}}

\newcommand{\LeadingB}[1]{%
    {\Ylinecolour{BrickRed}\gyoung(#1)}%
    }
    
\newcommand{\tworow}[1]{%
    {\Ylinecolour{ForestGreen}\gyoung(#1)}%
    }

\newcommand{\threerow}[1]{%
    {\Ylinecolour{ProcessBlue}\gyoung(#1)}%
    }

\newcommand{\wtwo}{\Yfillcolour{BrickRed}}
\newcommand{\wtwosec}{\Yfillcolour{ForestGreen}}
\newcommand{\wonesec}{\Yfillcolour{NavyBlue}}
\newcommand{\wfour}{\Yfillcolour{Orange}}
\newcommand{\wthree}{\Yfillcolour{Orchid}}
\newcommand{\wfive}{\Yfillcolour{GreenYellow}}

\title{An introduction to string states and their interactions}
\author*[a]{Chrysoula Markou}

\affiliation[a]{ Scuola Normale Superiore and INFN,\\
  Piazza dei Cavalieri 7, 56126 Pisa, Italy}

\emailAdd{chrysoula.markou@sns.it}

\abstract{The subject matter of these lecture notes is the open bosonic critical string, its perturbative spectrum and interactions. We begin with a brief review of classical string propagation, quantization, as well as the level--by--level construction of physical string states. We then review a new, covariant and efficient technology of constructing entire trajectories of infinitely many physical states deeper in the string spectrum. Finally, elements of the calculation of string scattering amplitudes, including aspects of the application of the technology for the efficient calculation of tree--level amplitudes of deeper trajectories, are also covered. The material is based on three invited lectures delivered by the author at the 2024 Modave Summer School in Mathematical Physics.}

\FullConference{Modave Summer School in Mathematical Physics (Modave2024)\\
 1 September 2024\\
Modave, Belgium\\}

\begin{document}
\maketitle

%%%%%%%%%%%%%%%%%%%%%%%%%%%%%%%%%%%%%%%%%%%%%%%%%%%%%%%%%%%%%
\section{Introduction}
\label{sec:intro}
%%%%%%%%%%%%%%%%%%%%%%%%%%%%%%%%%%%%%%%%%%%%%%%%%%%%%%%%%%%%%

Back in the 60's, hadron physics was at the center of interest of the high energy physics community. Experimental evidence had arranged hadrons in almost linear \textit{trajectories} in the plane formed by their spins and masses--squared, while no known field theory could reproduce the soft high--energy behavior of hadronic scattering. Soon enough, QCD was discovered as the underlying microscopic theory, but the observationally motivated hypothesis  of \textit{planar duality} of e.g. the $4$--pion scattering amplitude, namely that the same amplitude can be written equally well as a sum of $s$--channel or of $t$--channel poles, also led to another discovery in the intermediate years. In particular, with the goal of constructing a $4$--point massive scalar amplitude with planar duality,  Veneziano postulated the following formula for its $s$--channel (up to an overall prefactor) \cite{Veneziano:1968yb}
\begin{align} \label{eq:ven}
    \mathcal{A}_s(s,t) = \frac{\Gamma\big(\alpha(s) \big) \Gamma\big(\alpha(t)\big)}{\Gamma \big(\alpha(s) + \alpha(t) \big)}
\end{align}
and similar ones for the other channels, where $\Gamma$ is the gamma function, $s,t,u$ are the three Mandelstam variables and 
\begin{align}
    \alpha(s) := -\alpha' s -1 \,, \quad  \alpha(t) := -\alpha'  t -1 \,,
\end{align}
$\alpha'$ being a parameter with mass dimension $-2$. Using the properties of the gamma function, it can be shown that the \textit{Veneziano amplitude} enjoys planar duality, as well as very \textit{soft} UV behavior. Moreover, the formula \eqref{eq:ven} conceals a very rich structure: the amplitude has infinitely many \textit{simple poles} at
\begin{align}
    s = \frac{N-1}{\alpha'} \,,
\end{align}
similarly for the other channels, where $N$ is a an unbounded integer starting from $0$.

Very soon after, it was realized independently by Nambu \cite{Nambu:1969se}, Nielsen \cite{Nielsen:1970} and Susskind \cite{Susskind:1970xm} that the Veneziano amplitude was the 4--point amplitude of the lightest (yet tachyonic) state of a theory with an \textit{infinite} number of \textit{physical} states,  namely the theory of an open (bosonic) propagating string. Its closed string counterpart, formed by identifying the two open string endpoints, was also soon shown by Yoneya \cite{Yoneya:1973ca,Yoneya:1974jg} and Scherk and Schwarz \cite{Scherk:1974ca} to contain a single massless symmetric rank--$2$ tensor in its spectrum with its low--energy interactions matching precisely those of the graviton of General Relativity. On this note, and despite the absence of any experimental evidence to date, string theory remains an attractive area of research largely for two reasons: its spectrum contains the graviton and its scattering amplitudes, including gravitational ones, are free from UV divergences. At the origin of this last property lies precisely the \textit{infinity} of states of the perturbative string spectrum, the vast majority of which are massive \textit{higher--spin} string states. Consequently, decoding the string spectrum is crucial to decrypting string theory's role as framework towards a quantum theory of gravity and ultimately its fundamental underlying principles. 

Recently, an effort towards enhancing our understanding of the string spectrum was made with the development of a new, covariant and efficient technology of constructing entire trajectories of infinitely many physical string states and their interactions at one go \cite{Markou:2023ffh}. In $S$-matrix language, this amount to building algorithmically \textit{daughters} of a \textit{parent} trajectory, for example of the leading Regge trajectory, deeper in the spectrum, the presence of all of which being tied to the good UV behavior of string amplitudes. The daughters are constructed by means of a \textit{spectrum--generating} symplectic algebra. It is a symplectic algebra that had been hidden in the Virasoro constraints that impose physicality of string states and which commutes with the spacetime Lorentz algebra, (of the little group) of which the polarizations of all string states are irreducible representations, thereby enabling the use of a powerful result from representation theory known as Howe duality \cite{Howe1}--\cite{Howe2}. The formula for a generalized Koba--Nielsen factor, that can serve as the basic ingredient for the $N$--point scattering amplitude of \textit{any} trajectory, was also given in \cite{Markou:2023ffh}, including examples of several $3$--point amplitudes.

These lecture notes are meant to be a pedagogical and brief review of long known as well as more recently found results on the string spectrum and string scattering amplitudes, covering the three $2$--hour lectures delivered by the author at Modave2024. We begin by reviewing the rudiments of the classical propagation of the bosonic string, its action principles, quantization as well as the construction of the spectrum on a level--by--level basis in Sections \ref{sec:action}--\ref{sec:level}, following mainly the books \cite{Green:2012oqa,Polchinski:1998rq, Blumenhagen:2013fgp}. We have also especially benefited from Gleb Arutyunov's lecture notes \cite{Arutyunov}, as well as by the lectures notes compiled by Niklas Beisert and Johannes Br\"odel \cite{Beisert}, Oliver Schlotterer \cite{Schlotterer} and Timo Weigand \cite{Weigand}. Part of the presentation in Section \ref{sec:level} is motivated by the formalism of $\mathfrak{so}$ representations and their Young diagrams employed in the context of the new technology \cite{Markou:2023ffh}, to which we turn in Section \ref{sec:traj}. Elements of the calculation of string scattering amplitudes including the $3$--gluon and $3$--graviton amplitudes are given in Section \ref{sec:int}, while supplementary material for subleading trajectories from \cite{Markou:2023ffh}, that was not covered during the lectures, appears in Section \ref{sec:genf}. The list of references is by no means claimed to be complete and notification of typos and other issues spotted will be appreciated by the author.

%%%%%%%%%%%%%%%%%%%%%%%%%%%%%%%%%%%%%%%%%%%%%%%%%%%%%%%%%%%%%
\section{The string action}
\label{sec:action}
%%%%%%%%%%%%%%%%%%%%%%%%%%%%%%%%%%%%%%%%%%%%%%%%%%%%%%%%%%%%%

Let us imagine a string propagating in an ambient spacetime. The former is a $1$--dimensional object, which may be \textit{open} or \textit{closed}, and the latter say $D$--dimensional, referred to in the literature as the \textit{string background} or \textit{target space}. As the string moves in the background, it sweeps out a $2$--dimensional surface, the \textit{worldsheet} $\Sigma$. Let us parametrize it with the coordinates $\sigma^\alpha=(\tau, \sigma)$, where  $\alpha, \beta,\ldots =0,1$, with $\tau$ and $\sigma$ respectively denoting time and space on the worldsheet. If we further assume that the background is flat, namely associated with a metric $\eta_{\mu \nu}$ of Minkowski signature (we use the mostly plus convention), then the embedding of the worldsheet in the background may be defined by the map or \textit{embedding}
\begin{equation} \label{eq:map}
X^\mu(\tau, \sigma): \Sigma \rightarrow \mathbb{R}^{1,D-1}\,.
\end{equation}
For closed strings, periodicity of the map $X^\mu$ in the spacial coordinate $\sigma$ must be imposed, which may be written as the condition
\begin{equation} \label{eq:period}
X^\mu(\tau,\sigma+\ell)= X^\mu(\tau, \sigma)\,,
\end{equation}
where $\ell$ is the length of the string, conventionally taken as $\ell=2\pi$ in the literature and in the following.

What kind of action describes such a string propagation? A ``natural'' candidate proposed by Nambu \cite{Nambu:1970dh}, Goto \cite{Goto:1971ce} and Hara \cite{Hara:1971ur}, mirroring actions for point particles, is the total area the string sweeps out, 
\begin{equation} \label{eq:NG}
S_{\textrm{NG}} = - T\, \int_{\Sigma} \md A\,,
\end{equation}
where $dA$ is the Minkowski area element. Since its mass dimension is $-2$ (in natural units), it requires the introduction of a parameter $T$ with mass dimension $+2$ for a consistent action, whose interpretation we will think of in a bit. To write the action \eqref{eq:NG} as an integral of a function of $X^\mu$, it is instructive to first compute the Euclidean area element $\md A_{\textrm{Eucl}}$. For that, one first needs to perform a Wick rotation in spacetime, setting $X^0 :=t= \mathrm{i} w$, where $w$ is now a spatial coordinate and the embedding coordinates become
\begin{equation}
\vec{X}(\tau, \sigma): \Sigma \rightarrow \mathbb{R}^{D}\,.
\end{equation}
The Euclidean area element $\md A_{\textrm{Eucl}}$ is then simply the area of the parallelogram formed by the tangent vectors $\md \vec{\ell}_\sigma$ and $\md \vec{\ell}_\tau$ at any point of the worlsdheet $\Sigma$, namely the tangent vectors' cross product
\begin{equation}
\md A_{\textrm{Eucl}} = | \md \vec{\ell}_\sigma \times \md \vec{\ell}_\tau | = | \md \vec{\ell}_\sigma |\,|  \md \vec{\ell}_\tau | \, \sin{\theta}\,,
\end{equation}
where $\theta$ is the formed by the two tangent vectors. Using the notation
\begin{equation}
\dot{} : = \partial /\partial \tau \,, \quad ' : = \partial  /\partial \sigma\,, \quad \partial_\alpha := \frac{\partial}{ \partial \sigma^\alpha} \,, \quad \md^2 \sigma = \md \tau \md \sigma \,,
\end{equation}
one then has
\begin{equation}
\md A_{\textrm{Eucl}} = \md^2 \sigma \, \sqrt{\dot{\vec{X}}^2 {\vec{X}}^{'2} - \big(\dot{\vec{X}} \cdot \vec{X}'\big)^2}\,,
\end{equation}
where the dot product is to be thought of as the Euclidean scalar product. But let us look at the matrix
\begin{equation}
\gamma_{\alpha \beta} := \partial_\alpha X^\mu \partial_\beta X^\nu \eta_{\mu \nu} := \partial_\alpha X \cdot \partial_\beta X\,,
\end{equation}
where now the dot product stands for the Minkowski scalar product. Do you recognize this matrix? It is the induced metric from the ambient spacetime to the $2$--dimensional worldsheet. In Euclidean coordinates it takes the form
\begin{equation}
\gamma_{\alpha \beta}^{\textrm{Eucl}} =  \partial_\alpha \vec{X} \cdot \partial_\beta \vec{X}
\end{equation}
(where the dot product stands for the Euclidean scalar product), whose determinant reads
\begin{equation}
\det \gamma_{\alpha \beta}^{\textrm{Eucl}} :=  \gamma_{\textrm{Eucl}}  = \dot{\vec{X}}^2 {\vec{X}}^{'2} - \big(\dot{\vec{X}} \cdot \vec{X}'\big)^2\,,
\end{equation}
so we  learn that
\begin{equation}
\md A_{\textrm{Eucl}} = \md^2 \sigma \sqrt{ \gamma_{\textrm{Eucl}}}\,.
\end{equation}

Is the Minkowski analogue $\gamma$ of $ \gamma_{\textrm{Eucl}}$ positive--definite? Let us investigate, turning back to Minkowski spacetime with $X^0=t =\tau$ (the second identification being an allowed choice of coordinates) and $\vec{\upsilon}=d\vec{X}/dt$, where now $\vec X$ is a vector in $D-1$ dimensions. Then
\begin{align}
\dot{X}^2 & = -\Big(\frac{dt}{d\tau}\Big)^2 + \Big(\frac{d\vec{X}}{d\tau}\Big)^2 = -1+\vec{\upsilon}^2 \leq 0 \,, \\
X'^2 &= -\Big(\frac{dt}{d\sigma}\Big)^2 + \Big(\frac{d\vec{X}}{d\sigma}\Big)^2 = \Big(\frac{d\vec{X}}{d\sigma}\Big)^2 \geq 0\,, \\
\dot{X} \cdot X' &= - \frac{dt}{d\tau} \cdot \frac{dt}{d\sigma} + \frac{d\vec{X}}{d\tau} \frac{d\vec{X}}{d\sigma} = \vec{\upsilon} \cdot \frac{d\vec{X}}{d\sigma} \,,
\end{align}
so $\gamma$ does not have a definite sign in general. However, let us look at a generic linear combination of the two vectors $\dot{X}^\mu$, $X'^\mu$, namely a family of tangent vectors parametrized by $\lambda$
\begin{align}
S^\mu = \dot{X}^\mu + \lambda X'^\mu\,.
\end{align}
The norm of the vector $S^\mu$ \,,
\begin{align}
S_\mu S^\mu = \dot{X}^2 + 2\lambda\, \dot{X} \cdot X' + \lambda^2 X'^2\,,
\end{align}
we may view as a polynomial in $\lambda$ with discriminant
\begin{align}
\Delta = 4 \Big[ \big(\dot{X} \cdot X'\big)^2- \dot{X}^2 X'^2 \Big] = -4\gamma \,.
\end{align}
However, at every point on the worldsheet there must always be a unique timelike and a unique spacelike tangent vector, which requires $\Delta >0$, such that the equation $S_\mu S^\mu =0$ always have two solutions. Consequently, \textit{causal propagation} of the string requires $\gamma <0$, so the area element in Minkowski spacetime reads $\sqrt{-\gamma}$ and the action takes the form
\begin{align}
S_{\textrm{NG}} = -T \int_{\Sigma} \md^2 \sigma \sqrt{-\gamma}\,.
\end{align}

The presence of the square root famously renders the action hard to quantize. One solution is to construct a \textit{polynomial} action upon introducing an additional and auxiliary field on the worldsheet, the worldsheet's \textit{intrinsic} metric $h_{\alpha \beta}(\tau,\sigma)$ (with Minkowski signature). Such an action was constructed by two groups, Brink, Di Vecchia and P.~S.~Howe \cite{Brink:1976sc} and Deser and Zumino \cite{Deser:1976rb} and is known as the Polyakov action. It takes the form
\begin{align}\label{actionP}
S_{\textrm{P}} = -\tfrac{T}{2} \int_\Sigma \md^2 \sigma \sqrt{-h} h^{\alpha \beta} \gamma_{\alpha \beta} = -\tfrac{T}{2} \int_\Sigma \md^2 \sigma  \sqrt{-h} \tr(\gamma)
\end{align}
and includes two metrics, the induced $\gamma_{\alpha \beta}$, which contains the dynamical field $X^\mu$, and the auxiliary $h_{\alpha \beta}$, on which no derivatives act, consistently with it being non--dynamical. The action $S_{\textrm{P}}$ thus describes the coupling of $D$ massless scalars, namely the components of $X^\mu$, to gravity in two dimensions! Now is also an appropriate point to reflect on the nature of the parameter $T$. Recall its mass dimension, $[T]=2\,$: in natural units, $[\textrm{time}]= [\textrm{length}]=-1$, $[\textrm{mass}]=1$ and $[\textrm{force}]=2$, so we learn that $T$ is a constant force (namely constant energy/unit length)! This implies that a string is \textit{neither a spring} (recall Hook's law) \textit{nor a rubber band}. The string \textit{tension} $T$ is in fact \textit{the only free} (and dimensionful) parameter in string theory. In the literature it's often parametrized as
\begin{align}
T=\frac{1}{2\pi \alpha'}\,,
\end{align}
where $\alpha$ is known as the Regge slope and is precisely the dimensionful parameter that appears in the Veneziano amplitude \eqref{eq:ven} and to which we will turn back at a later point.

The equations of motion (e.\ o.\ m.\ ) of the fields $X^\mu$ and $h_{\alpha \beta}$ derived from the action \eqref{actionP} read (Exercise\footnote{ Hint: use that $\delta h = - h\, h_{\alpha \beta} \delta h^{\alpha \beta}$, which is easy to see by using Jacobi's formula for an invertible matrix $\mathbf{A}(\rho)$ depending on a variable $\rho$, namely $\frac{d}{d\rho} \det \mathbf{A}(\rho) = \det \mathbf{A}(\rho) \tr(\mathbf{A}^{-1}(\rho) \frac{d}{d\rho} \mathbf{A}(\rho)) = - \det \mathbf{A}(\rho) \tr[ \Big(\frac{d}{d\rho} \mathbf{A}^{-1}(\rho) \Big) \mathbf{A}(\rho) ]  \,$.}) 
\begin{align}\label{eom1}
\Box X^\mu & := \frac{1}{\sqrt{-h}} \partial_\alpha \big(\sqrt{-h} h^{\alpha \beta} \partial_{\beta} X^\mu) \overset{!}{=}0 \\ \label{eom2}
T_{\alpha \beta} & := \frac{4\pi}{\sqrt{-h}} \frac{\delta S_{\textrm{P}}}{\delta h^{\alpha \beta}} = - \frac{1}{\alpha'} \Big( \partial_\alpha X \cdot \partial_\beta X - \tfrac12 h_{\alpha \beta} h^{\gamma \delta} \partial_\gamma X \cdot \partial_\delta X \Big) \overset{!}{=}0\,,
\end{align}
where $T_{\alpha \beta}$ is the energy--momentum tensor on the worldsheet. The components of $h_{\alpha \beta}$ enter the action \eqref{actionP} as Lagrange multipliers, so their e.\ o.\ m.\ \eqref{eom2} impose \textit{constraints} on the kinematics of $X^\mu$. Upon solving \eqref{eom2} for $h$ and substituting in the action \eqref{actionP}, it is easy to see that the actions $S_{\textrm{P}}$ and $S_{\textrm{NG}}$ are classically equivalent (Exercise\footnote{Hint: use that $\det(c \mathbf{A})= c^n \det(\mathbf{A})$ for a $n \times n$ matrix $\mathbf{A}$ and a parameter $c$.}). Let us also note that for the open string (of length $\pi$), the boundary condition, 
\begin{align} \label{eq:openbc}
\int_{\partial \Sigma} \md n^\alpha \partial_\alpha X \cdot \delta X = 0\,,
\end{align}
where $n^\alpha$ is the normal vector at every point of the worldsheet, has further to be imposed to arrive at \eqref{eom1}; this is easy to see by applying Stokes' theorem when varying the action.

Let us now consider the symmetries of the action $S_{\textrm{P}}$. These fall into two categories:
\begin{enumerate}
\item global: Poincar\'e transformations in spacetime, with respect to which $X^\mu $ is vector and $h_{\alpha \beta}$ a scalar, since $X$ appears only within $\partial X$.
\item local:
\begin{enumerate}
\item worldsheet diffeomorphisms (diffeos), which infinitesimally read $\sigma^\alpha \rightarrow \sigma^\alpha +\xi^a(\sigma,\tau)$, under which $X^\mu$ transforms as a scalar and $h_{\alpha \beta}$ as a tensor, namely (Exercise)
\begin{align}
\delta X^\mu = - \xi^\alpha \partial_\alpha X^\mu \,,\quad \delta h_{\alpha \beta} =- (\nabla_\alpha \xi_\beta + \nabla_\beta \xi_\alpha) \,,
\end{align}
where $\nabla_\alpha$ is the covariant derivative on the worldsheet (for a torsionless metric $h_{\alpha \beta}$). Moreover, this symmetry implies the \textit{conservation} of the worldsheet energy--momentum tensor (Exercise)
\begin{align}
\nabla^\alpha T_{\alpha \beta} =0\,.
\end{align}
\item Weyl rescalings of the worldsheet intrinsic metric, which infinitesimally read $\delta h_{\alpha \beta}= 2\Lambda(\tau,\sigma) \,h_{\alpha \beta}$  (while of course $\delta X^\mu=0$), which imply the \textit{tracelessness} of the worldsheet energy--momentum tensor (Exercise)\,,
\begin{align}
T^\alpha_{\hphantom{\alpha }\alpha} := h^{\alpha \beta} T_{\alpha \beta}=0\,.
\end{align}
\end{enumerate}
\end{enumerate}

It turns out to be particularly useful to gauge--fix part of the worldsheet redundancies. Recall that the metric $h_{\alpha \beta}$ is a symmetric rank--$2$ tensor and as such has $\tfrac12 d(d+1)=3$ independent components (where $d=2$ stands for the number of worldsheet dimensions). Reparametrization invariance is associated with $2$ parameters, namely the components of  $\xi_\alpha$, so it can be used to fix $2$ components of the intrinsic metric and bring it to a conformally flat form
\begin{align}
h_{\alpha \beta}(\sigma) = f(\sigma) \eta_{\alpha \beta}\,.
\end{align}
Weyl rescalings are associated with one parameter $\Lambda$, so also the third component of the metric can be fixed, e.g. by setting $f(\sigma)=1$. The metric becomes then
\begin{align}
h_{\alpha \beta}^{\textrm{c.g.}} = \eta_{\alpha \beta}\,,
\end{align}
which is known as the \textit{conformal gauge} and to which we will be restricting ourselves from now on. This is a famous result: in two dimensions, the graviton can be gauged away \textit{completely}. In this gauge, the action takes the form
\begin{align}
S_{\textrm{P}}^{\textrm{c.g.}}= -\tfrac{T}{2} \int_\Sigma \md^2 \sigma \,  \eta^{\alpha \beta} \partial_\alpha X \cdot \partial_\beta X\,,
\end{align}
which is invariant under a residual symmetry, namely under those diffeos that can be undone by Weyl rescalings: this is the so--called \textit{conformal} symmetry of the worldsheet. The e.\ o.\ m.\ of $h_{\alpha \beta}$ become 
\begin{align} \label{Vir1}
T_{00}&=-\frac{1}{2\alpha'} \big(\dot{X}^2 + X'^2 \big)=T_{11} \overset{!}{=}0 \\ \label{Vir2}
T_{01}&=-\frac{1}{\alpha'} \dot{X} \cdot{X}'=T_{10} \overset{!}{=}0,
\end{align}
which are known as the (classical) \textit{Virasoro constraints}. The constraint \eqref{Vir2} in particular implies that the two vectors tangent to the worldsheet are always perpendicular to each other, namely that the string moves in a direction that is perpendicular to the direction it is stretched out in: the string has \textit{no inner structure}! The e.\ o.\ m.\ of $X^\mu$ takes the form
\begin{align} \label{eq:eomxcg}
\partial^2 X^\mu = - \ddot{X}^\mu + X''^\mu \overset{!}{=} 0\,,
\end{align}
which is the massless harmonic wave equation in two dimensions.

%%%%%%%%%%%%%%%%%%%%%%%%%%%%%%%%%%%%%%%%%%%%%%%%%%%%%%%%%%%%%
\section{Solving the e.\ o.\ m.\ and the space of states}
\label{sec:eom}
%%%%%%%%%%%%%%%%%%%%%%%%%%%%%%%%%%%%%%%%%%%%%%%%%%%%%%%%%%%%%

To solve the e.\ o.\ m.\ of  $X^\mu$, it is practical to choose light cone coordinates on the worldsheet,
\begin{align}
\sigma^{\pm}:= \tau \pm \sigma \,,\quad \partial_{\pm}:=\frac{\partial}{\partial\sigma^{\pm}}=\tfrac12(\partial_\tau\pm \partial_\sigma) \,,\quad  \partial^2 := \partial_\alpha \partial^\alpha = -4\partial_+ \partial_-\, \,,
\end{align}
in which the general solution of \eqref{eq:eomxcg} takes the form
\begin{align}
X^\mu(\tau,\sigma)=X^\mu_{\textrm{L}} (\sigma^+)+X^\mu_{\textrm{R}}(\sigma^-)\,,
\end{align}
namely $X^\mu$ can be written as a sum of two fields, denoted by the indices L (left) and R (right), that depend on $\sigma^+$ and $\sigma^-$ respectively. The Virasoro constraints \eqref{Vir1}--\eqref{Vir2} take the form (Exercise)
\begin{align}\label{virnew}
(\partial_+ X)^2 = 0= (\partial_- X)^2\,.
\end{align}
Let us distinguish then the following cases.

(i) Closed strings. Since in this case $X^\mu$ periodic in $2\pi$, the discrete Fourier expansions of $X^\mu_{\textrm{L}}$ and $X^\mu_{\textrm{R}}$ read\footnote{The rescaling $\sigma \rightarrow \sigma'  = \frac{\ell}{2\pi} \sigma$ recovers the correct mass dimensions.}
\begin{align} \label{Fourier1}
X^\mu_{\textrm{L}} (\sigma^+)& = \tfrac12 x^\mu + \tfrac12 \alpha' p^\mu \sigma^+ + \mathrm{i} \sqrt{\tfrac{\alpha'}{2}} \sum_{n\neq 0} \frac{1}{n} \, \tilde{\alpha}_n^\mu \, e^{-  \mathrm{i} n\sigma^+} \,, \\ \label{Fourier2}
X^\mu_{\textrm{R}} (\sigma^-) &= \tfrac12 x^\mu + \tfrac12 \alpha' p^\mu \sigma^- + \mathrm{i} \sqrt{\tfrac{\alpha'}{2}} \sum_{n\neq 0} \frac{1}{n} \, \alpha_n^\mu \, e^{-  \mathrm{i} n\sigma^-} \,,
\end{align}
where $x^\mu$, $p^\mu$ are parameters (that are constant w.\ r.\ t.\ the worldsheet coordinates) and  the parameters $ \tilde{\alpha}_n^\mu\,$, $ \alpha_n^\mu\,$, $n\in\mathbb{Z}^*$, are the Fourier modes of the string field $X^\mu$, frequently also called \textit{string modes} or \textit{oscillators} or L and R \textit{movers} respectively. Because of the reality of $X^\mu$, the parameters $x^\mu$, $p^\mu$ are real and (Exercise)
\begin{align}
\alpha_{-n}^\mu = (\alpha_n^\mu)^* \,,\quad \tilde{\alpha}_{-n}^\mu = (\tilde{\alpha}_n^\mu)^*\,,
\end{align}
and the prefactors in \eqref{Fourier1}--\eqref{Fourier2} are chosen for later convenience. The parameter $p^\mu$ enters in such away that $X^\mu \big|_{\alpha , \tilde{\alpha}= 0} = x^\mu + \alpha' p^\mu \tau $ respects the periodicity condition. In fact, as we will see later, the pair $(x^\mu, p^\mu)$ is describes the string's center--of--mass motion, with $x^\mu$ corresponding to its position and $p^\mu$ to its momentum.

Next, the Virasoro constraints have to be imposed on the solution \eqref{Fourier1}--\eqref{Fourier2}. First, we have that
\begin{align}
\partial_+ X^\mu = \partial_+ X_{\textrm{L}}^\mu = \sqrt{\tfrac{\alpha'}{2}} \sum_{n\in \mathbb{Z}}\tilde{\alpha}_n^\mu \, e^{- \mathrm{i} n\sigma^+}\,,\quad
\partial_- X^\mu = \partial_- X_{\textrm{R}}^\mu = \sqrt{\tfrac{\alpha'}{2}} \sum_{n\in \mathbb{Z}} \alpha_n^\mu \, e^{- \mathrm{i} n\sigma^-}\,,
\end{align}
where we have introduce the $0$--mode oscillators by setting
\begin{align}
\tilde{\alpha}_0^\mu := \sqrt{\tfrac{\alpha'}{2}}p^\mu =: \alpha_0^\mu\,.
\end{align}
Consequently, we have that
\begin{align}
(\partial_+ X)^2 = \tfrac{\alpha'}{2} \sum_{n,\ell \in \mathbb{Z}} \tilde{\alpha}_n \cdot \tilde{\alpha}_\ell \, e^{- \mathrm{i} (n+\ell)\sigma^+} = \alpha' \sum_{m\in \mathbb{Z}}\Big( \tfrac12 \sum_{\ell \in \mathbb{Z}} \tilde{\alpha}_{m-\ell} \cdot \tilde{\alpha}_\ell \Big) e^{- \mathrm{i} m\sigma^+}
\end{align}
(similarly for $(\partial_- X)^2$) and by defining the \textit{Virasoro modes}
\begin{align} \label{eq:Vir_modes}
L_n  := \tfrac12 \sum_{m \in \mathbb{Z}} \alpha_{n-m} \cdot \alpha_m \,,\quad \tilde{L}_n = \tfrac12 \sum_{m \in \mathbb{Z}} \tilde{\alpha}_{n-m} \cdot \tilde{\alpha}_m\,,
\end{align}
which can be shown to be the Fourier modes of the worldsheet energy--momentum tensor $T_{\alpha \beta}$ (Exercise), the (classical) Virasoro constraints \eqref{Vir1}--\eqref{Vir2} become
\begin{align}
\sum_{n\in \mathbb{Z}} L_n e^{- \mathrm{i}  n \sigma^+} \overset{!}{=} 0 \overset{!}{=}  \sum_{n\in \mathbb{Z}} \tilde{L}_n e^{- \mathrm{i} n \sigma^-} \quad \Rightarrow \quad L_n = 0 =\tilde{L}_n \quad \forall n \in \mathbb{Z}\,. 
\end{align}
We have thus derived that \textit{all} Virasoro modes have to vanish classically. 

(ii) Open strings. As we have seen,  the boundary condition \eqref{eq:openbc} has to be imposed. This we may rewrite as
\begin{align}\label{eq:bc2}
\int_0^\pi \md \sigma \, \partial^\tau X \cdot \delta X \big|_{\tau_i}^{\tau_f} + \int_{\tau_i}^{\tau_f} \md \tau \, \partial^\sigma X \cdot \delta X \big|_0^\pi \overset{!}{=} 0\,,
\end{align}
where $(\tau_i, 0)$ and $(\tau_f, \pi)$ are the string's initial and final conditions respectively, with $\partial^\tau X = - \dot{X} \,, \partial^\sigma X = X' $. The first term in the expression \eqref{eq:bc2} vanishes assuming that
\begin{align}
\delta X |_{\tau_i} = \delta X |_{\tau_f} =0\,, 
\end{align}
namely that the initial and final conditions are specified. Two ways of having then the second term vanish are by imposing the
\begin{enumerate}
\item \label{it:N} Neumann (N) boundary conditions:
\begin{align}
 X'^\mu |_{\sigma = 0,\pi} =0\,,
\end{align}
which implies that there is no momentum flow off the ends of the string, or the
\item \label{it:D}  Dirichlet (D) boundary conditions:
\begin{align}
\delta X^\mu |_{\sigma = 0,\pi} = 0\,,
\end{align}
which fixes the spacetime positions of the string endpoints, thereby breaking spacetime Poincar\'e invariance. The string's spacetime momentum is thus not conserved and the string endpoints are thought of as being confined on hypersurfaces that are also dynamical objects known as D--\textit{branes}.
\end{enumerate}
The conditions \ref{it:N} and \ref{it:D} can be chosen independently for every direction $\mu$. In these lectures we will restrict ourselves to \textit{spacetime--filling} D--branes, namely impose N conditions for both string endpoints in all directions, so that spacetime Poincar\'e invariance is unbroken. 

Moreover, there appears now only one set of Fourier modes of the field $X^\mu$, which may intuitively be understood due to a reflection of L into R modes (or vice versa) at the string boundary, namely due to the identification
\begin{align} \label{eq:ident}
\alpha_n^\mu \equiv \tilde{\alpha}_n^\mu
\end{align}
and similarly there is only one set of Virasoro modes $L_n$. The open string expansion of the field $X^\mu$ thus reads
\begin{align}
X_{\textrm{o}}^\mu (\tau,\sigma) =  x^\mu + 2\alpha' p^\mu \tau + \mathrm{i} \sqrt{2 \alpha'} \sum_{n\neq 0} \frac{\alpha^\mu_n}{n} e^{-\mathrm{i} n\tau} \cos(n \sigma)\,,
\end{align}
where now we have defined 
\begin{align} \label{eq:azero_open}
\alpha_0^\mu := \sqrt{2\alpha'} p^\mu\,. 
\end{align}
The normalization in the expansion is such that the center--of--mass position and momentum of \textit{both} the closed and the open string are identified with the pair $(x^\mu,p^\mu)$; more specifically, $p^\mu$ is the Noether charge of (global) spacetime translations in both cases (Exercise). Finally, the open string Virasoro constraints take the form 
\begin{align} \label{eq:Vir_class}
L_n = 0  \,, \quad \forall  n \in \mathbb{Z}\,.
\end{align}
It is worth mentioning that open string expressions can be obtained from the closed string counterparts by performing the so--called \textit{doubling trick}, which loosely speaking amounts to performing the identification \eqref{eq:ident} as well as the replacement
\begin{align} \label{eq:doubling}
p^\mu \rightarrow 2 p^\mu \,, \quad X_{\textrm{L}}^\mu \rightarrow \tfrac12 X_{\textrm{o}}^\mu \,, \quad X_{\textrm{R}}^\mu \rightarrow \tfrac12 X_{\textrm{o}}^\mu \,,
\end{align}
since a closed string can be thought of as two open strings with identified endpoints, each carrying half the closed string's momentum.

We are now ready to review the old canonical quantization of the bosonic string in the conformal gauge. The canonical momentum is classically defined as
\begin{align}
\Pi^\mu := \frac{\partial \mathcal L_{\textrm{P}}^{\textrm{c.g.}}}{\partial \dot{X}^\mu} = T \dot{X}^\mu
\end{align}
and one may impose equal $\tau$ time Poisson brackets
\begin{align}
\{X^\mu(\tau, \sigma),X^\nu(\tau, \sigma')\}_{\textrm{P.B.}}  &= 0 = \{\dot{X}^\mu(\tau, \sigma),\dot{X}^\nu(\tau, \sigma')\}_{\textrm{P.B.}}  \\
\{X^\mu(\tau, \sigma), \dot{X}^\nu(\tau, \sigma')\}_{\textrm{P.B.}} & = \tfrac{1}{T} \,\eta^{\mu \nu } \delta(\sigma-\sigma') \,.
\end{align}
Quantum mechanically, one may consider $X^\mu$ as well as $x^\mu,p^\mu, \alpha_n^\mu\,\tilde{\alpha}_n^\mu\,$, as quantum mechanical operators, so the standard replacement
\begin{align}
\{\,,\,\}_{\textrm{P.B.}}  \rightarrow \tfrac{1}{\mathrm{i}} [\,,\,]
\end{align}
yields the equal time commutators (Exercise)
\begin{align}
[x^\mu,p^\nu] &= \mathrm{i} \eta^{\mu \nu} \,,\quad [x^\mu , x^\nu] =0 = [p^\mu,p^\nu] \\ \label{eq:osc_alg}
[\alpha_m^\mu,\alpha_n^\nu] &= m\delta_{m+n} \eta^{\mu \nu} = [\tilde{\alpha}_m^\mu,\tilde{\alpha}_n^\nu]\,,\quad [\alpha_m^\mu,\tilde{\alpha}_n^\nu] = 0\,.
\end{align}
The L and R sectors of the closed string are thus coupled only via $x^\mu$ and $p^\mu$ (and as discussed in the open string there exists only one copy of modes $\alpha_n^\mu$). Consequently, after rescaling, the quantization of the bosonic string yields \textrm{infinitely} many \textit{independent} copies of the algebra
\begin{align}
[a,a^\dagger] =1\,,
\end{align}
namely of the quantum \textit{harmonic oscillator} (namely infinitely many independent copies in the L and in the R sectors of the closed string). Finally, note that reality now becomes hermiticity:
\begin{align}
(\alpha_n^\mu)^\dagger = \alpha_{-n}^\mu\,.
\end{align}

It is thus possible to construct a Fock space of string states as usual. Restricting to open strings, a number operator,
\begin{align}
N_m := \,:\alpha_m \cdot \alpha_{-m}: ( =\alpha_{-m} \cdot \alpha_m) \,, \quad m>0\,,
\end{align}
may be defined to this end, which acts on the oscillators as
\begin{align}
[N_m, \alpha_m] = -m\, \alpha_m \,,\quad [N_m,\alpha_{-m}] = +m\, \alpha_{-m} \,,\quad \forall \, m>0\,.
\end{align}
For $m>0$, $\alpha_m$ and $\alpha_{-m}$ may hence be thought of as annihilation and creation modes respectively and normal ordering, denoted above by the symbol ``$:\,:$'', imposes that all annihilation operators are placed to the right of all creation operators within a normal--ordered expression. The vacuum $\ket{0;p^\mu}$, namely the ground state of the Fock space, is then defined via
\begin{align}
\alpha_m^\mu \ket{0;p^\mu} = 0 = \tilde{\alpha}_m^\mu \ket{0;p^\mu} \quad (m>0)\,,\quad \hat{p}^\mu \ket{0;p^\mu} = p^\mu \ket{0;p^\mu} \,,
\end{align}
namely it is annihilated by all annihilation operators and, $\hat{p}^\mu$ being the quantum momentum operator, it is a momentum eigenstate, so it can be written as a plane--wave times the zero--momentum ground state,
\begin{align}
\ket{0;p^\mu} = e^{\mathrm{i} p\cdot X} \ket{0}\,.
\end{align}
Open string state candidates can then be built out of creation operators $\alpha_{-n}^\mu$, $n>0$, acting on the vacuum, with their spacetime indices contracted with an a priori arbitrary tensor of $GL(D)$, such that they be spacetime scalars. Such a state candidate thus takes the form
\begin{align} \label{eq:cand}
\ket{\textrm{candidate}}= \varepsilon_{\mu \nu \lambda \dots}  \alpha_{-i}^\mu \alpha_{-j}^\nu \alpha_{-k}^\lambda \dots \ket{0;p} \,, \quad i,j,k,\ldots >0 \,,
\end{align}
so we already have evidence that the string spectrum contains \textit{higher--spin} states! However, upon considering, for example, the object $\alpha_{-n}^\mu \ket{0}$ and its norm,
\begin{align} \label{eq:testn}
\bra{0} \alpha_n^\mu \alpha_{-n}^\mu  \ket{0} = n \, \eta^{\mu \mu}\,,
\end{align}
we notice that the latter becomes negative for $\mu=0$. This means that a priori the spectrum may contain \textit{negative--norm states}, that are pathological ``ghosts'', which would pose a problem for the probabilistic interpretation of the quantized string.

This problem is resolved by restricting to the physical space of states, namely the subspace that obeys the Virasoro constraints. Recalling that classically they take the form \eqref{eq:Vir_class}, namely all Virasoro modes, defined in \eqref{eq:Vir_modes}, have to vanish, let us first consider whether the quantum version of the Virasoro modes $L_n$ can be defined simply by normal--ordering their classical definition. For $L_{n \neq 0}$, there is no ambiguity, as $\alpha_{n-m}$  and $\alpha_m$ always commute in this case, so
\begin{align} \label{eq:vir_def}
L_{n \neq 0} := \tfrac12 \sum_{m \in \mathbb{Z}}  :\alpha_{n-m} \cdot \alpha_m: \,.
\end{align}
On the other hand, since $[\alpha_{-m}^\mu,\alpha_m^\nu]=-m \eta^{\mu \nu}$, two definitions are a priori possible for $L_0$, namely
\begin{align} \label{eq:L0}
L_0 := \tfrac12 \alpha_0^2 + \sum_{m=1}^\infty \alpha_{-m} \cdot \alpha_m + D \sum_{m=1}^\infty m  \quad \textrm{or} \quad L_0 := \tfrac12 \alpha_0^2 + \sum_{m=1}^\infty \alpha_{-m} \cdot \alpha_m\,,
\end{align}
so there is an ambiguity up to the \textit{normal ordering constant}
\begin{align}
a := D \sum_{m=1}^\infty m \,,
\end{align}
which, crucially, can be regularized to a number. We may then \textit{choose} the second of definitions \eqref{eq:L0} and replace $L_0 \rightarrow L_0 + a$ in the quantum version of the Virasoro constraints, which we now turn to.

%%%%%%%%%%%%%%%%%%%%%%%%%%%%%%%%%%%%%%%%%%%%%%%%%%%%%%%%%%%%%
\section{The Virasoro constraints and physical states}
\label{sec:con}
%%%%%%%%%%%%%%%%%%%%%%%%%%%%%%%%%%%%%%%%%%%%%%%%%%%%%%%%%%%%%
Given that the classical Virasoro constraints \eqref{eq:Vir_class} imply that physical states are those that are annihilated by \textit{all} Virasoro modes, it appears enticing to write their quantum version as
\begin{align} \label{eq:Vir_quant1}
(L_0-a) \ket{\textrm{phys}} =0  = L_n \ket{\textrm{phys}}  \,,\quad \forall \, n\in \mathbb{Z}^*\,,
\end{align}
at first glance. However, in the quantum string, the Virasoro operators $L_n$ satisfy the famous infinite--dimensional \textit{Virasoro algebra} 
\begin{align} \label{eq:Vir_alg}
[L_m,L_n] = (m-n)L_{m+n} + \tfrac{D}{12} \, m \, (m^2-1)  \, \delta_{m+n}\,,
\end{align}
which can be derived by using the oscillator algebra \eqref{eq:osc_alg} (Exercise). Without the second term in the RHS of \eqref{eq:Vir_alg}, the algebra is known as the \textit{Witt algebra}, which in the context of the string was discovered by Fubini and Veneziano \cite{Fubini:1971ce}, to which the second term is thought of as a \textit{central extension}, as  found by J.\ H.\ Weis in unpublished work\footnote{We thank Paolo Di Vecchia for a clarification of this point.} (see for example \cite{DiVecchia:2007vd}). The number $D$ of spacetime dimensions here thus also plays the role of the central charge of the worldsheet conformal field theory. Now let us imagine that we impose the constraints \eqref{eq:Vir_quant1} on a physical state $\ket{\textrm{phys}}$. We would then have that
\begin{align}
\langle \textrm{phys} | [L_n,L_{-n}] | \textrm{phys} \rangle =0 \quad \Rightarrow \quad  \big[ 2n \, a + \tfrac{D}{12} \, n \, (n^2-1)  \big] \langle \textrm{phys} |  \textrm{phys} \rangle  =0\,,
\end{align}
so, for the last equality to hold for any physical state of non--zero norm, we would obtain
\begin{align}
 2n \, a + \tfrac{D}{12} \, n \, (n^2-1) = 0\,,
\end{align}
which cannot hold simultaneously $\forall n \in \mathbb{Z}^*$. This is a signal of the presence of the \textit{Weyl anomaly} at the quantum level: the Virasoro constraints, being first--class constraints at the classical level as they are associated with Weyl invariance, become \textit{second--class} at the quantum level as Weyl invariance is broken, unless additional elements are introduced (this is best seen in the path integral quantization of the string, which we do not review in these lectures). To solve this problem, we may impose that only a subset of the Virasoro modes annihilate physical states, namely that
\begin{align} \label{eq:physicality}
(L_0-a) \ket{\textrm{phys}} =0  = L_{n>0} \ket{\textrm{phys}}  \,,
\end{align}
for example. In fact, an even smaller number of constraints is \textit{sufficient} to construct all physical states, since due to the Virasoro algebra
\begin{align}
[L_{n+1},L_1] = n L_{n+2}\,, \quad n >0\,,
\end{align}
so annihilation of a state by $L_1$ and $L_2$ implies annihilation by all other positive Virasoro modes. In other words, only \textit{three} constraints are sufficient to construct the physical states of the open string,
\begin{align} \label{eq:Vir_suff}
(L_0-a) \ket{\textrm{phys}} =0  \,,\quad  L_1 \ket{\textrm{phys}}  =0\,, \quad L_2 \ket{\textrm{phys}}  =0  \,.
\end{align}

What about the mass of the physical states? Recall that for open strings $\alpha_0^\mu$ is related to the string's center--of--mass total momentum $p^\mu$ via \eqref{eq:azero_open}. Squaring and using the definition of the mode $L_0$ and the respective Virasoro constraint according to the discussion after \eqref{eq:L0} yields
\begin{align} \label{eq:mass}
M^2 = \frac{N-a}{\alpha'}\,,
\end{align}
where we have defined the level $N$ via
\begin{align}
N :=  \sum_{m=1}^\infty N_m = \sum_{m=1}^\infty \alpha_{-m} \cdot \alpha_m\,,
\end{align}
which is clearly an unbounded non--negative integer (when acting on string states). Consequently, \eqref{eq:mass} yields the mass spectrum of the open string: it is an \textit{infinite tower} of physical string states, organized in mass levels $N=0,1,2,\dots$. Importantly, since $\alpha'$ sets the mass scale of \textit{all} states, there is \textit{no parametric scale separation} between string states that find themselves at different levels. Constructing the physical states in the \textit{old covariant} way amounts then to writing an Ansatz constructed out of all (combinations of) creation oscillators that may be excited at the desired level $N$ and imposing the Virasoro constraints \eqref{eq:Vir_suff} thereon. In this way, and as we will see soon in explicit examples, the a priori arbitrary tensors $\varepsilon_{\mu \nu \lambda \dots}$ that enter the polynomials of candidate states \eqref{eq:cand} become the states' polarization tensors and, as such, they must correspond to irreducible representations of the little group of the spacetime Lorentz group  $SO(D-1,1)$ that is generated by the bilinears (Exercise)
\begin{align} \label{eq:lorentzgen}
J^{\mu \nu } = 2 p^{[\mu} x^{\nu]} -2 \mathrm{i} \sum_{n=1}^\infty \frac{1}{n}  \alpha_{-n}^{[\mu} \alpha_n^{\nu]}\,,
\end{align}
the little group being $SO(D-1)$ and $SO(D-2)$ for massive and massless states respectively (Exercise\footnote{A helpful review in this regard can be \cite{Bekaert:2006py}.}).

The parameters $a$ and $D$ have been left arbitrary until now, but in fact their values can be precisely determined and there are several ways of doing so. We will now start building the lightest levels of the spectrum, which will also turn out to be perhaps the simplest way of finding allowed values for $a$ and $D$. At the lightest level, $N=0$, so no oscillator can be excited and there appears only one state that is a scalar with mass $M^2= -\frac{a}{\alpha'}$, so at this point we cannot say whether it is massless or massive (and perhaps tachyonic). At the next level, $N=1$, so only the first oscillator can be excited and the Ansatz reads
\begin{align} \label{eq:vec1}
\ket{\textrm{candidate}} = \varepsilon \cdot \alpha_{-1} \ket{0;p}\,,
\end{align}
where $\varepsilon^\mu=\varepsilon^\mu(p)$ is an a priori arbitrary vector of $GL(D)$. Indeed, the action of the level operator $N$, which essentially counts the number of oscillators excited to produce a given state, on the first oscillator yields
\begin{align}
N \alpha_{-1}^\mu \ket{0;p} = \alpha_{-1}\cdot \alpha_1 \,\alpha^\mu_{-1} \ket{0;p}=\alpha_{-1}^\mu \ket{0;p}.
\end{align}
The mass of the vector state is given by $M^2=\frac{1}{\alpha'}(1-a)$, while its norm reads
\begin{align}
    \bra{0;p} \varepsilon \cdot \alpha_1 \, \varepsilon \cdot \alpha_{-1} \ket{0;p} = \varepsilon^2\,,
\end{align}
so at this point one may not exclude that it even be a negative--norm state, as alluded to with \eqref{eq:testn}.

Towards excluding this possibility, let us consider \textit{spurious} states, a class of states that the student reader may already be familiar with from a quantum field theory course, the QFT here being the \textit{conformal field theory} associated with the field $X^\mu$ from the point of view of the worldsheet, which we will turn to slightly more thoroughly in Section \ref{sec:int}. Spurious are states that are orthogonal to all physical states and thus decouple from physical processes, namely string interactions here. Spurious states may have zero or negative norm: the former are both physical and orthogonal to all other states and are called \textit{null} states, while the latter cannot be accepted as physical states. Null states thus satisfy the mass--shell and the physicality conditions 
\begin{align}
    (L_0-a) \ket{\textrm{null}} =0\,, \quad L_{n>0} \ket{\textrm{null}}  = 0\,,
\end{align}
as well as orthogonality 
\begin{align}
    \bra{\textrm{phys}} \ket{\textrm{null}} = 0\,.
\end{align}
Consequently, the simplest way to construct spurious states is to have the \textit{negative} Virasoro modes act on some states $\ket{\chi_n}$ as in
\begin{align} \label{eq:null}
\ket{\textrm{null}}  = \sum_{n=1}^\infty L_{-n} \ket{\chi_n}\,.
\end{align}
Checking the on--shell condition on the null state \eqref{eq:null} and using the Virasoro algebra \eqref{eq:Vir_alg} imposes a type of on--shell condition on the state $\ket{\chi_n}$,
\begin{align} \label{eq:L0chi}
 \,(L_0 -a +n) \ket{\chi_n} =0 \,.
\end{align}
Moreover, again due to the Virasoro algebra, the Virasoro modes $L_{-1} $ and $L_{-2}$ can generate all other negative Virasoro modes, so the states
\begin{align}
\Big\{  L_{-1} \ket{\chi_1} ,\,  L_{-2} \ket{\chi_2} \Big\}
\end{align}
are a sufficient set of ingredients to construct all null states.  Upon a factor of proportionality, the two simplest null states then read 
\begin{align} \label{eq:null1}
\ket{\textrm{null}}_1 & \sim L_{-1} \ket{\chi_1} \\ \label{eq:null2}
\ket{\textrm{null}}_2 & \sim \Big( L_{-2}  +  \gamma L_{-1}^2 \Big) \ket{\chi_2}\,,
\end{align}
where $\gamma$ is a parameter. If there exist such states in the open bosonic string, we would have
\begin{align}
L_1 \ket{\textrm{null}}_1 \overset{!}{=}0  \,,\quad L_2 \ket{\textrm{null}}_2 \overset{!}{=} 0 \quad \Rightarrow \quad a=1\,, \quad D=26\,,
\end{align}
and also
\begin{align}
L_1 \ket{\textrm{null}}_2 \overset{!}{=} 0 \quad \Rightarrow \quad \gamma=3/2\,.
\end{align}

With this information, let us look again at the vector \eqref{eq:vec1}. For $a=1$ we have that
\begin{align}
p^2=-M^2 =0\,, 
\end{align}
so it is a massless state. Let us check the Virasoro constraints on the vector. Using the definition \eqref{eq:vir_def} of $L_{n\neq 0}$, the action of $L_1$ on \eqref{eq:vec1} yields
\begin{align}
\big(\alpha_0 \cdot \alpha_1 + \alpha_{-1} \cdot \alpha_2 + \alpha_{-2}\cdot \alpha_3 + \dots \big) \, \varepsilon\cdot \alpha_{-1} \ket{0;p} \overset{!}{=}0 \quad \Rightarrow \quad \alpha_0 \cdot \varepsilon = 0 \,,
\end{align} 
so, using the definition \eqref{eq:azero_open}, one finds
\begin{align}
p\cdot \varepsilon = 0 \,,
\end{align}
namely that the state's polarization is transverse to its momentum. Since it is massless, we may choose the rest frame $p^\mu =(E,E,0,\dots,0)$, in which the polarization takes the form $\varepsilon^\mu = (0,0, \vec \varepsilon)$, so one further finds that the state's norm is positive, since $\varepsilon^2= |\vec \varepsilon|^2$. Furthermore, the action of $L_2$  on \eqref{eq:vec1} yields
\begin{align}
\big(\alpha_1 \cdot \alpha_1 + \alpha_0 \cdot \alpha_2 + \alpha_{-1}\cdot \alpha_3 + \dots \big) \, \varepsilon\cdot \alpha_{-1} \ket{0;p} \overset{!}{=}0 \,,
\end{align} 
which is trivially satisfied.

Now let us consider the null states at level $N=1$. There exists only one, 
\begin{align}
\ket{\textrm{null}} = \tfrac{\mathrm{i}}{\sqrt{2\alpha'}} \Lambda \, L_{-1} \ket{\chi_1} \,, \quad  (L_0-1) \ket{\textrm{null}}=0 \,,\quad p^2 = 0\,,
\end{align}
where $\Lambda$ is an arbitrary parameter (with the prefactor chosen for convenience) and the state $\ket{\chi_1}$ finds itself at a certain level $\tilde N$. The condition \eqref{eq:L0chi} then yields
\begin{align}
L_0 \ket{\chi_1} = 0 \quad \Rightarrow  \quad \tilde N\ket{\chi_1}  = 0 \quad \Rightarrow \quad \ket{\chi_1} = \ket{0;p}\,. 
\end{align}
Consequently, using the definition \eqref{eq:vir_def} for $L_{-1}$, we find that
\begin{align} \label{eq:nullN1}
\ket{\textrm{null}} = \tfrac{\mathrm{i}}{2\sqrt{2\alpha'}} \big( \alpha_{-1} \cdot \alpha_0 + \alpha_0 \cdot \alpha_{-1}  + \dots \big) \ket{0;p} = \mathrm{i} \Lambda\, p \cdot \, \alpha_{-1} \ket{0;p}
\end{align}
since all terms in the dots contain annihilation operators, each multiplied with a commuting creation operator. What have we learned? While the vector's polarization $\varepsilon^\mu$ is in general a function of the momentum $p^\mu$, the null state \eqref{eq:nullN1}  encodes the degrees of freedom of a \textit{longitudinally polarized} vector, whose polarization enjoys transversality precisely because $\ket{\textrm{null}}$ is massless, namely
\begin{align}
    i \Lambda p \cdot p = 0\,,
\end{align}
so it is physical but null. The states
\begin{align}
    \varepsilon \cdot \alpha_{-1} \ket{0;p} \quad \textrm{and} \quad \big(\varepsilon_\mu +i \Lambda \, p_\mu\big)\, \alpha_{-1}^\mu \ket{0;p}
\end{align}
thus describe the same physics and so may be identified, which can be expressed in terms of the appearance of a \textit{gauge symmetry},
\begin{align}
    \varepsilon^\mu \rightarrow \varepsilon^\mu + i \Lambda \, p^\mu\,.
\end{align}
Consequently, the physical state at level $N=1$ has $24$ propagating d.\ o.\ f.\ in $D=26$, as two components of $\varepsilon^\mu$ can be removed by the constraint of transverality and by a gauge fixing respectively, so that it becomes a vector of $SO(D-2)$.

%%%%%%%%%%%%%%%%%%%%%%%%%%%%%%%%%%%%%%%%%%%%%%%%%%%%%%%%%%%%%
\section{On the level--by--level construction of the spectrum}
\label{sec:level}
%%%%%%%%%%%%%%%%%%%%%%%%%%%%%%%%%%%%%%%%%%%%%%%%%%%%%%%%%%%%%

So far we have learned that null states exist in the open string spectrum for $a=1$ and $D=26$. It can further be shown that \textit{negative--norm} states decouple for $D \leq 26$ (with $a=1$), with the $D=26$ string often called the \textit{critical} string. The absence of negative--norm states for $D = 26$ is hard to prove in the old covariant quantization, as a check of unitarity at the $1$--loop level is needed. In the \textit{light cone gauge} (which we will briefly touch upon later on), the absence of ghosts for $D \leq 26$ was shown by Brower \cite{Brower:1972wj} and by Goddard and Thorn \cite{Goddard:1972iy}, in what is known as the ``no--ghost'' theorem. But let us look at the first few levels of the critical string. At $N=0$, it is now clear that the scalar has mass--squared $M^2 = -\frac{1}{\alpha'}$ and is thus a \textit{tachyon} of $1$ d.\ o.\ f.\ The appearance of the tachyon is a sign of instability of the vacuum, implying that the true vacuum of the bosonic string is another, stable, state and is an interesting problem per se, with significant progress in the calculation of the open string tachyon potential having been made by Sen and Zwiebach \cite{Sen:1999nx}\footnote{The proofs of the no--ghost theorem, the treatment of the tachyon and the superstring, in which it is possible to project the tachyon out of the spectrum, are outside the scope of the present introductory lectures.}.  At $N=1$, as discussed, there appears again only one state, 
\begin{align}\label{eq:masslessv}
    \varepsilon \cdot \alpha_{-1} \ket{0;p} \,,\quad \varepsilon \cdot p = 0 \,, \quad p^2 =0 \,,
\end{align}
namely the transverse massless vector with $24$ propagating d.\ o.\ f.\ ; this is the only massless state in the entire open string spectrum. At $N=2$, the Ansatz for its state content reads
\begin{align}
    \ket{\textrm{candidate}} = \Big( \tilde{\varepsilon} \cdot \alpha_{-2} + \varepsilon_{\mu \nu} \alpha_{-1}^\mu \alpha_{-1}^\nu \Big) \ket{0;p}\,, \quad M^2 = - p^2 = \tfrac{1}{\alpha'}\,,
\end{align}
where $ \tilde{\varepsilon}^\mu$ and $\varepsilon^{\mu \nu}$ are two a priori arbitrary tensors of $GL(D)$ (with $\varepsilon^{\mu \nu}$ being symmetric by construction). We then have that (Exercise)
\begin{align} \label{eq:N2const}
\begin{aligned}
    L_1 \ket{\textrm{candidate}} &\overset{!}{=}  0 \quad \Rightarrow \quad \sqrt{2\alpha'} p^\mu \varepsilon_{\mu \nu} + \varepsilon_\nu = 0  \\
    L_2 \ket{\textrm{candidate}} &\overset{!}{=}  0 \quad \Rightarrow \quad \varepsilon^\mu{}_\mu + 2 \sqrt{2\alpha'}  p \cdot \varepsilon = 0 \,,
\end{aligned}
\end{align}
where $ \varepsilon^\mu{}_\mu = \varepsilon_{\mu \nu} \eta^{\mu \nu}$ is the trace of the rank--2 tensor $\varepsilon_{\mu \nu}$. 

Let us consider the null states at $N=2$; they are two, namely
\begin{align}
    \ket{\textrm{null}}_1 = \Lambda_1 L_{-1} \ket{\chi_1} \,, \quad    \ket{\textrm{null}}_2 = \Lambda_2 \Big( L_{-2} +\tfrac32  L_{-1}^2 \Big) \ket{\chi_2}\,,
\end{align}
where $\ket{\chi_1}$ and $\ket{\chi_2}$ are at this point unknown states at levels $N_1$ and $N_2$ respectively and $\Lambda_1$, $\Lambda_2$ are arbitrary parameters. Due to the constraint \eqref{eq:L0chi}, we then have that
\begin{align}
\begin{aligned}
        L_0 \ket{\chi_1} & \overset{!}{=} 0  \quad \Rightarrow \quad N_1 = 1 \quad \Rightarrow \quad \ket{\chi_1}  = \xi \cdot \alpha_{-1} \ket{0;p}\\
     (L_0+1) \ket{\chi_2} & \overset{!}{=} 0 \quad \Rightarrow \quad N_2 = 0 \quad \Rightarrow \quad \ket{\chi_2} = \ket{0;p}\,,
\end{aligned}
\end{align}
where $\xi$ is an a priori arbitrary vector. Using the definition \eqref{eq:vir_def} for $L_{-1}$ and $L_{-2}\,$ (and imposing physicality of the null states), we can then deduce that (Exercise)
\begin{align}
    \begin{aligned}
         \ket{\textrm{null}}_1 & =  \Lambda_1 \big( \xi \cdot \alpha_{-2} + \sqrt{2\alpha'} \, p\cdot  \alpha_{-1} \, \xi \cdot \alpha_{-1} \big) \ket{0;p} \,,\quad  p \cdot \xi = 0 \,,\\
         \ket{\textrm{null}}_2 & = \Lambda_2 \Big[ \tfrac52 \sqrt{2\alpha'} \, p \cdot \alpha_{-2} + \tfrac12 \,\alpha_{-1} \cdot \alpha_{-1} +3\alpha' (p \cdot \alpha_{-1})^2 \Big] \ket{0;p}\,.
    \end{aligned}
\end{align}
Consequently, we may think of the transverse and longitudinal components of $\tilde{\varepsilon}_\mu$ as being encoded in  $\ket{\textrm{null}}_1$ and in $ \ket{\textrm{null}}_2 $ respectively, while the rest of the terms in $\ket{\textrm{null}}_1$ and in $ \ket{\textrm{null}}_2 $ correspond to the longitudinal d.\ o.\ f.\ of $\varepsilon_{\mu \nu}$ (for example its trace is encoded in $ \ket{\textrm{null}}_2 $). Equivalently, we may think of the two null states as each being associated with a type of ``St\"uckelberg symmetry'' of the level's state content, namely that the physical state content, equivalently well the set of constraints \eqref{eq:N2const}, is invariant under
\begin{align}
\begin{aligned}
  \tilde{\varepsilon}_\mu  &\rightarrow \tilde{\varepsilon}_\mu + \Lambda_1 \xi_\mu  \,,\quad   \varepsilon_{\mu \nu} &\rightarrow  \varepsilon_{\mu \nu} +\sqrt{2\alpha'} \Lambda_1  p_{(\mu} \xi_{\nu)} 
\end{aligned}
\end{align}
and (independently) under
\begin{align}
\begin{aligned}
  \tilde{\varepsilon}_\mu  &\rightarrow \tilde{\varepsilon}_\mu + \tfrac52 \Lambda_2  \sqrt{2\alpha'} p_\mu  \,,\quad   \varepsilon_{\mu \nu} &\rightarrow  \varepsilon_{\mu \nu}  + \tfrac12 \Lambda_2 \, \eta_{\mu \nu}  + 3\alpha' \Lambda_2  \, p_\mu p_\nu
\end{aligned}
\end{align}
(noticing that the transformation of $\varepsilon_{\mu \nu}$ is ``traceless'' and ``tracefull'' under each symmetry respectively). We may thus use these redundancies to gauge away $\tilde{\varepsilon}^\mu$ completely, so, using the constraints \eqref{eq:N2const} we find that the level $N=2$ contains a single massive state,
\begin{align} \label{eq:lightestspin2}
     \varepsilon_{\mu \nu} \alpha_{-1}^\mu \alpha_{-1}^\nu  \ket{0;p} \,,\quad  p^\mu \varepsilon_{\mu \nu}  = 0  = \varepsilon^\mu{}_\mu  \,,\quad p^2 = -\tfrac{1}{\alpha'} 
\end{align}
that is a transverse and traceless (TT) massive spin--$2$ state, with $\varepsilon_{\mu \nu}$ now an irreducible tensor of $SO(D-1)$. It propagates $\frac12 D (D+1) - D - 1 = \tfrac12 (D+1)(D-2)$ d.\ o. f.\ (after subtracting the number of TT constraints), namely $324$ d.\ o.\ f.\ in $D=26$.

Further comments are in order. Beyond $N=2$, every level contains at least $2$ states and always a finite number thereof, with the density of states given by the \textit{Hardy--Ramanujan formula} in the asymptotic limit $N \rightarrow \infty$. Apart from the massless vector, all other states are massive and the vast majority are \textrm{higher--spin} states, all of which are physical and TT, with the null states encoding the longitudinal components including traces. The entire spectrum consists of on-shell TT states only, which can be constructed on a level--by--level basis in the old--covariant way; while this method is algorithmic, clearly its application increases in difficulty as the level increases, since the Ans\"atze become longer and solving the Virasoro constraints, which are complicated and non--linear, hence harder. An alternative is to use the residual conformal symmetry of the worldsheet (after choosing the conformal gauge) to make a non--covariant gauge choice, the previously referred to \textit{light cone gauge}  \cite{Goddard:1973qh}. More specifically, light cone coordinates are chosen on the worldsheet as in
\begin{align}
X^\pm = \tfrac{1}{\sqrt 2} \big(X^0 \pm X^{D-1} \big)
\end{align}
and then a gauge in which $X^+$ is linear in $\tau$ and has no oscillators is chosen. The only free creation oscillators are then the transverse oscillators $\alpha_{-n}^i$, $i=1,\dots, D-2$, so any function of $\alpha_{-n}^i$ is automatically physical, without needing to solve the Virasoro constraints. However, the output is given in terms of $SO(D-2)$ irreps, \textit{recombining} which into $SO(D-1)$ irreps also becomes harder and harder as the level increases. A similar limitation is faced by the powerful formalism developed by Del Giudice, Di Vecchia and Fubini (DDF) \cite{DelGiudice:1971yjh} that is adapted to string scattering amplitudes; the formalism is not manifestly covariant because of the presence of a reference momentum in the parametrization of physical polarizations. 

Other than that, all open string states must further be dressed by the prefactor $g_{\textrm{o}} T^a$. The parameter $g_{\textrm{o}}$ is the \textit{open string coupling}, which sets the strength of open string interactions and ensures the correct mass dimension of string scattering amplitudes. It is not free, but rather set by the vacuum expectation value of the dilaton $\Phi$, which is a massless state of the closed string spectrum, via $g_{\textrm{o}} \sim e^{\langle \Phi \rangle /2}$. More specifically, the string coupling $g_{\textrm{S}}$ can be traced back to the term 
\begin{align} \label{eq:Euler}
  \tfrac{\lambda}{4 \pi} \int_\Sigma \md^2 \sigma \sqrt{-h} R = \lambda \chi\,,
\end{align}
where $\lambda$ is a parameter and $R$ the worldsheet Ricci scalar. The term \eqref{eq:Euler} may be added to the Polyakov action \eqref{actionP} (but breaks Weyl rescaling invariance) and $\lambda$ turns out to be the dilaton's vev $\langle \Phi \rangle$ and $\chi$ is a \textit{topological invariant} known as the Euler characteristic of the surface $\Sigma$, which here is the string worldsheet. For $2$--dimensional surfaces like the worldsheet, the integrand in \eqref{eq:Euler} is a total derivative, so it does not affect the classical e.\ o.\ m.\, but does play a role in the quantum string in perturbation theory. In particular, its contribution to the path integral is 
\begin{align}
    e^{\langle \Phi \rangle \chi} := g_{\textrm{S}}^\chi\,,
\end{align}
so that an expansion in the string coupling $g_{\textrm{S}}$ is an expansion in string worldsheet topology.  The couplings $g_{\textrm{S}}$ and $g_{\textrm{o}}$ are thus dynamically determined and not input; they are proportional to each other and the proportionality constant is a power of $\alpha'$ (the only free parameter), that can be determined by considering string scattering. Turning to the $T^a$, as previously discussed, the open string endpoints are constrained to lie on a D--brane, but we may further think of a stack of $\mathcal{N}$ ``coincident'' D--branes, different branes among which the string endpoints may find themselves attached to. In that case, by associating labels $k,\ell=1,\dots \mathcal{N}$, namely the so--called Chan--Paton labels with every open string state, we see that are $\mathcal{N}^2$ configurations for every physical string state. These we may package in $\mathcal{N} \times \mathcal{N}$ matrices $T^a$, the \textit{Chan--Paton factors}, that are $\mathcal{N}^2$ in number. One can also show that, for string gauge interactions to be consistent, the matrices $T^a$ have to be hermitian, so they can be seen as the generators of the group $U(\mathcal{N})$  \cite{Paton:1969je}. In other words, in the case of one brane only, there exists a massless abelian vector in the string spectrum that is associated with the group $U(1)$, while for $\mathcal{N}$ branes the vector is non--abelian, associated with $U(\mathcal{N})$ and turns out to be described by Yang--Mills theory at low energies, a point that we will come back to in Section \ref{sec:int}. 

So what do heavier (open) string states, the presence of \textit{all} of which is crucial to the good UV behavior of string scattering amplitudes, look like? What does the \textit{deep} string spectrum look like?  We have learned that any massive physical state can be represented by a suitable function $ F$ of the creation operators $\alpha_{-n}^\mu$, 
\begin{align}
    F(\alpha_{-1}^\mu,\alpha_{-2}^\nu,\dots) \ket{0;p}\,,
\end{align}
where the oscillators' indices are contracted with the state's TT polarization tensor (that is $p$--transverse w.\ r.\ t.\ every index and traceless w.\ r.\ t.\ any pair of indices), which is an irreducible tensor of $\mathfrak{so}(D-1)$. With a view of soon turning to the technology of \cite{Markou:2023ffh} for the construction of entire \textit{trajectories} of infinitely many physical states, let us first gather a few key properties of irreducible tensors\footnote{Appendix E of \cite{Didenko:2014dwa} is a concise and helpful review of irreducible tensors in this regard.} that were used in \cite{Markou:2023ffh}. To construct a massive state's polarization, we may start with an irreducible tensor of $\mathfrak{gl}(D)$, imposing tracelessness on which renders it an irrep of $\mathfrak{so}(D)$, which in turn becomes an irrep of $\mathfrak{so}(D-1)$ after imposing transversality. A general irreducible tensor of $\mathfrak{gl}(D)$ can take the form $\varepsilon^{\mu(s_1),\, \lambda (s_2), \dots,\, \nu(s_L)}\,(p) $, where the notation $\mu(s_1) := \mu_1 \dots \mu_{s_1} $ stands for a group of $s_1$ symmetric indices $\mu$, separated from other symmetric groups by a comma, which may be represented by a Young diagram in the \textit{symmetric base} as 
\begin{align} \label{eq:yds}
 \Yboxdim{14pt} \gyoung(_9{s_1},_8{s_2},_6{\dots},_4{s_L}) \quad  \Leftrightarrow \quad \varepsilon^{\mu(s_1),\, \lambda (s_2), \dots,\, \nu(s_L)}\,(p) \,.
\end{align}
In this picture, every index of the polarization tensor is represented by a box of the Young diagram, with the group of $s_i$ symmetric indices corresponding to a row of $s_i$ boxes that hence has length $s_i$.

The tensor \eqref{eq:yds} is then irreducible iff it enjoys ``Young symmetry,'' namely if symmetrization of all indices of a group $k$ with one index of any subsequent group $\ell$ is zero. Let us consider the example of the ``hook'' $ \gyoung(;;,;) \,\, \Leftrightarrow \,\,  \varepsilon^{\mu_1 \mu_2,\lambda} $. Its Young symmetry condition takes the form 
\begin{align}
\varepsilon^{\mu_1 \mu_2,\lambda} + \varepsilon^{ \mu_2  \lambda , \mu_1} + \varepsilon^{\lambda \mu_1, \mu_2} =0 \quad \textrm{or} \quad \textrm{``}\varepsilon^{\mu \mu, \mu} =0 \textrm{''} \,,
\end{align}
which essentially forces that it be linearly independent from the (irreducible) rank--$3$ symmetric tensor $ \gyoung(;;;)\,$. More generally, the Young symmetry condition in this notation can be written as 
\begin{align} \label{eq:Ysym}
    \varepsilon_{\ldots, \mu_k(s_k),\ldots, \mu_k \mu_\ell(s_\ell-1),\ldots} = 0 \,, \quad \textrm{for} \quad k<\ell\,,
\end{align}
so the lengths of rows can never increase as the number of rows increases, namely $s_1 \geq  s_2 \geq \ldots \geq s_n$ (otherwise the tensor vanishes).
To turn then the tensor \eqref{eq:yds} into an irrep of $\mathfrak{so}(D-1)$, the tracelessness condition,
\begin{align} \label{eq:traceless}
 \eta^{\rho \sigma}   \varepsilon_{\ldots, \rho\mu_k(s_k-1),\ldots, \sigma \mu_\ell(s_\ell-1),\ldots} = 0  \quad \textrm{and} \quad  \eta^{\rho \sigma}   \varepsilon_{\ldots, \rho \sigma \mu_k(s_k-2),\ldots} = 0 \,,
\end{align}
as well as the transversality condition 
\begin{align}
    p^\nu \varepsilon_{\ldots, \nu \mu_k (s_k-1),\ldots} = 0 \,,
\end{align}
must further be imposed. Let us note that it is not necessary to impose these conditions independently for any pair of indices or for any index to construct irreducible tensors; rather, a subset thereof is sufficient, with the rest following as a consequence, as we will see in an example in a bit. In the following, we will also restrict to the transverse subspace, which is sufficient to construct physical states \cite{Kato:1982im,Henneaux:1986kp, Manes:1988gz}, via
\begin{align} \label{trsub}
    \eta^{\mu \nu}_\perp=\eta^{\mu \nu} - \frac{p^\mu p^\nu}{p^2}\,,
\end{align}
namely restrict to $(D-1)$--spacetime dimensions and consider all states' polarizations as being already transverse, namely assume that they do not explicitly depend on the momentum $p^\mu$. This removes all longitudinal d.\ o.\ f.\  \textit{but the traces}, so we will have to impose the tracelessness condition \eqref{eq:traceless}.

In this notation, the first few levels of the open string are simply
\begin{align}
    \begin{aligned}
        N&=0:\quad \bullet \\
        N&=1: \quad  \gyoung(;)  \,,\quad F = \varepsilon \cdot \alpha_{-1} \\ 
        N&=2: \quad \gyoung(;;)\,,\quad F = \varepsilon_{\mu_1 \mu_2} \alpha_{-1}^{\mu_1} \alpha_{-1}^{\mu_2} \,,\quad \varepsilon^\mu{}_\mu = 0\,.
    \end{aligned}
\end{align}
As a side remark, the light cone states at $N=2$ are
\begin{align}
\alpha_{-1}^i \alpha_{-1}^ j \ket{0;p} \,,\quad \alpha_{-2}^i \ket{0;p}\,,
\end{align}
so a priori they correspond to the diagrams $\gyoung(;;)$ and $\gyoung(;)$ of $\mathfrak{gl}(D-2)$ respectively. After subtracting the trace of the former, the two light cone states yield the following  $\mathfrak{so}(D-2)$ irreps
\begin{align}
\gyoung(;;) \oplus \gyoung(;) \oplus \bullet\,,
\end{align}
which are simply recombined in the tensor $\gyoung(;;) $ of $\mathfrak{so}(D-1)$. However, as the level increases, more and more physical states appear per level, so recombination clearly becomes harder and harder to perform.

Turning back to the covariant way, what about a general string state? Clearly, a given Young diagram can be embedded in \textit{infinitely many different} ways in the string spectrum, by contracting the respective polarization in infinitely many different ways with the creation operators such that the Virasoro constraints be satisfied; these will embed the diagram at different mass levels, up to multiplicity. So a first question is, what is the \textit{simplest} way of contracting an $\mathfrak{so}(D-1)$ irreducible tensor with the $\alpha_{-n}^\mu$'s, such that the respective diagram is embedded at the \textit{lowest} possible level? The answer is to contract all boxes of the $i$--th row with the oscillator $\alpha_{-i}^\mu$ \cite{Weinberg:1985tv}, namely
\begin{align} \label{simplest}
    F_{\textrm{simpl}} = \varepsilon_{\mu(s_1),\, \lambda (s_2), \dots,\, \nu(s_L)} \, \alpha_{-1}^{\mu_1} \dots \alpha_{-1}^{\mu_{s_1}} \,\alpha_{-2}^{\lambda_1} \dots \alpha_{-2}^{\lambda_{s_2}}  \dots \alpha_{-L}^{\nu_1} \dots \alpha_{-L}^{\nu_{s_L}} \,,
\end{align}
which finds itself at 
\begin{align} \label{eq:lowestN}
 N_{\text{min}}=\sum_{i=1}^L s_i \,i 
\end{align}
and the states with polynomials of the type \eqref{simplest} are known as ``Weinberg states''. Due to the form \eqref{simplest}, rows are automatically symmetric and the level is the lowest possible, while we must further impose the Young symmetry \eqref{eq:Ysym}  and tracelessness condition \eqref{eq:traceless} (assuming also the transverse subspace simplification). Let us consider the example of $2$--row diagrams, 
\begin{align} \label{simplest2}
 \Yboxdim{10pt}  \gyoung(_4{s_1},_3{s_2})\,,\quad   F_{\textrm{simpl}} = \varepsilon_{\mu(s_1),\, \lambda (s_2)} \, \alpha_{-1}^{\mu_1} \dots \alpha_{-1}^{\mu_{s_1}} \,\alpha_{-2}^{\lambda_1} \dots \alpha_{-2}^{\lambda_{s_2}}   \,.
\end{align}
We then have that\footnote{The operator expressions are to be thought of as acting on the vacuum $\ket{0;p}$.}
\begin{align}
\begin{aligned} \label{eq:ltworow}
L_1 F_{\textrm{simpl}} & \overset{!}{=}0 \quad \Leftrightarrow \quad \big( \alpha_{-1} \cdot \alpha_2 + \dots \big)  F_{\textrm{simpl}} \overset{!}{=}0 \quad \Leftrightarrow \quad \varepsilon_{\mu(s_1),\mu \lambda(s_2-1)} = 0 \,, \\ 
L_2 F_{\textrm{simpl}} & \overset{!}{=}0 \quad \Leftrightarrow \quad \big( \alpha_1 \cdot \alpha_1 + \dots \big)  F_{\textrm{simpl}} \overset{!}{=}0 \quad \Leftrightarrow \quad \eta^{\rho \sigma} \varepsilon_{\rho \sigma \mu(s_1-2), \lambda(s_2)} = 0 \,, 
\end{aligned}
\end{align} 
so we see that $L_1$ and $L_2$ are satisfied \textit{iff} the $2$--row tensor obeys Young symmetry, i.e. total symmetrization of the first row with any of the (symmetric) indices of the second row yields zero, and is traceless w.\ r.\ t.\ to any pair of indices of the first row. Since there are no more than $2$ rows, there is no other relation imposed by Young symmetry, but what about the rest of the tracelessness conditions? Let us notice that
\begin{align}
L_3 F_{\textrm{simpl}} & \overset{!}{=}0 \quad \Leftrightarrow \quad \big( \alpha_1 \cdot \alpha_2 + \dots \big)  F_{\textrm{simpl}} \overset{!}{=}0 \quad \Leftrightarrow \quad \eta^{\rho \sigma} \varepsilon_{\rho \mu(s_1-1),\sigma \lambda(s_2-1)} = 0\,, 
\end{align} 
so we learn that every other trace is zero as a \textit{consequence} of the trace w.\ r.\ t.\ to any pair of indices of the first row being zero, since the $L_3$ constraint is automatically satisfied once the constraints \eqref{eq:ltworow} are satisfied and similarly with $L_4$ for the second row. Likewise, for Young diagrams with more rows, the Young symmetry condition w.\ r.\ t.\ any pair of rows is a consequence of a subset thereof.

In fact, it is possible \cite{Markou:2023ffh}  to rewrite the Young symmetry and tracelessness conditions via the action of an \textit{operator} $T^k{}_{\ell}$ that swaps $\alpha_{-\ell}$ with $\alpha_{-k}$ and of an operator $T_{k\ell}$ that takes a trace among the indices contracted with $\alpha_{-k}$ and $\alpha_{-\ell}$ respectively on $F_{\textrm{simpl}}$, namely by defining 
\begin{align} \label{eq:opT1}
    T^k{}_{\ell}  := \tfrac1k \, \alpha_{-k} \cdot  \alpha_{\ell} \, , \quad  T_{k\ell} :=  \alpha_k \cdot \alpha_{\ell} \,.
\end{align}
We then have that
\begin{align}
\begin{aligned}
 T^{k<\ell}{}_{\ell}  \,F_{\textrm{simpl}} & = 0 \quad \Leftrightarrow \quad  \varepsilon_{\ldots, \mu_k(s_k),\ldots, \mu_k \mu_\ell(s_\ell-1),\ldots} = 0 \\
 T_{k\ell} \, F_{\textrm{simpl}}  & = 0 \quad \Leftrightarrow \quad  \eta^{\rho \sigma}   \varepsilon_{\ldots, \rho\mu_k(s_k-1),\ldots, \sigma \mu_\ell(s_\ell-1),\ldots} = 0  
\end{aligned} 
\end{align}
(with annihilation by $T_{kk}$ forcing traceslessness w.\ r.\ t.\ a pair of indices of the same row), so the operators $ T^{k<\ell}{}_{\ell} $ and $ T_{k\ell}$ \textit{impose} (or check) Young symmetry and tracelessness on Weinberg states! For the $2$--row example \eqref{simplest2}, we thus have that (Exercise)
\begin{align}
\begin{aligned}
 T^1{}_2  \,F_{\textrm{simpl}} & = 0 \quad \Leftrightarrow \quad   \varepsilon_{\mu(s_1),\mu \lambda(s_2-1)}= 0 \\
 T_{11} \, F_{\textrm{simpl}}  & = 0 \quad \Leftrightarrow \quad  \eta^{\rho \sigma}   \varepsilon_{\rho \sigma \mu(s_1-2), \lambda(s_2)}  = 0\,,
\end{aligned} 
\end{align}
with annihilation by $T_{12}$ and $T_{22}$ following as consequences (and with all other  operators $T^1{}_3\,, T^2{}_3\,, \ldots$  and $ T_{22}\,, T_{23}\,,\ldots $ trivially annihilating $F_{\textrm{simpl}}$).

%%%%%%%%%%%%%%%%%%%%%%%%%%%%%%%%%%%%%%%%%%%%%%%%%%%%%%%%%%%%%
\section{Excavating string trajectories}
\label{sec:traj}
%%%%%%%%%%%%%%%%%%%%%%%%%%%%%%%%%%%%%%%%%%%%%%%%%%%%%%%%%%%%%

In this section, we review the main elements of the technology developed in \cite{Markou:2023ffh}\footnote{The technology was developed in the CFT language and without but also with the transverse subspace restriction; for pedagogical reasons we present it here in the equivalent oscillator language and with the transverse subspace simplification.}. We have seen that Weinberg states correspond to the embeddings of \textrm{all} possible $\mathfrak{so}(D-1)$ Young diagrams at the \textit{lowest} possible level, with the respective polynomials constructed with a simple recipe for the contraction with string oscillators. Yet what do more complicated states look like? To attempt to answer such questions, let us consider the physicality constraints \eqref{eq:Vir_suff} (for $a=1$, $D=26$) acting on a \textit{general} function $\ket{\textrm{phys}}$ of the creation operators $\alpha_{-1}^\mu\,,\alpha_{-2}^\nu\,,\ldots\,$. In the transverse subspace simplification, we may ignore the oscillator $\alpha_0^\lambda$ within the Virasoro modes $L_n$ (as well as the explicit dependence of the tensors in $\ket{\textrm{phys}}$ on the momentum $p$), so we have
\begin{align}\begin{aligned}
2 L_{n \neq 0} \ket{\textrm{phys}} & = \sum_{m=1}^\infty \big(  :\alpha_{n-m} \cdot \alpha_m: + :\alpha_{n+m} \cdot \alpha_{-m}: \big)\ket{\textrm{phys}}  \\
    & = \sum_{m=1}^\infty \big(  \alpha_{n-m} \cdot \alpha_m + \alpha_{-m}  \cdot  \alpha_{n+m} \big)\ket{\textrm{phys}} \\
    & = \bigg[ \sum_{m=1}^{n-1}   \alpha_{n-m} \cdot \alpha_m + \sum_{m=n+1}^\infty  \alpha_{n-m} \cdot \alpha_m + \sum_{m=1}^\infty \  \alpha_{-m}  \cdot  \alpha_{n+m} \bigg] \ket{\textrm{phys}}  \\
    & = \bigg[  \displaystyle \sum_{m=1}^{n-1} \alpha_{n-m}\cdot \alpha_m +2  \sum_{m=1}^\infty \alpha_{-m} \cdot \alpha_{n+m} \bigg]  \ket{\textrm{phys}} \,.
\end{aligned}
\end{align}
What do we observe in the last expression? In the first sum, we always have a product of two annihilation operators, while in the second the product (carrying $-n$ units of energy) of a creation with an annihilation operator. In other words, the two summands are the operators $T_{n-m,m}$ and $T^m{}_{n+m}$ respectively, namely the operators that check tracelessness and Young symmetry on $F_{\textrm{simpl}}$, as discussed in the previous section! We may thus rewrite the Virasoro constraints on a general state $\ket{\textrm{phys}}$ as
\begin{align} \label{eq:virfin}
   (L_0-1) \ket{\textrm{phys}} &= \bigg[ \displaystyle \sum_{m=1}^{\infty} nT^n{}_n + \alpha'p^2 -1\bigg]  \ket{\textrm{phys}} =0 \\ \label{eq:virfin2}
   2L_{n>0}  \ket{\textrm{phys}}& =\bigg[ \displaystyle \sum_{m=1}^{n-1}  \,T_{m,n-m} +2 \sum_{m=1}^\infty mT^m{}_{n+m} \bigg]  \ket{\textrm{phys}} =0\,.
\end{align}
With this reformulation, we see that Weinberg states pass the $L_1$ and $L_2$ constraints by construction, while the on--shell condition $L_0$ simply embeds them at the correct mass level \eqref{eq:lowestN}, with
\begin{align}
    T^n{}_n F_{\textrm{simpl}} \ket{0;p} = s_n F_{\textrm{simpl}} \ket{0;p} \,,
\end{align}
namely the operator $ T^n{}_n$ measures the length of the $n$--th row when it acts on a Weinberg state. But is it perhaps now possible to \textit{solve} the Virasoro constraints in an efficient way for more complicated sets of states?

Let us consider again the first few levels of the (open string) spectrum and in particular the state content in terms of the respective (little group) polarizations that is shown in table \ref{table:two}. The states whose Young diagrams have a red, green or light blue outline are all Weinberg states by definition. In fact, the states in red are the lightest member--states of a well--known \textit{trajectory} of states, the \textit{leading} Regge trajectory, see for example \cite{Sagnotti:2010at}. It consists of the highest--spin states per level, namely contains the spin--$s$ state at level $s$. In the language used here, it contains all first appearances of symmetric traceless (and transverse) $1$--row  tensors and so is described by a polynomial that we may write as
\begin{align} \label{leading}
F_{\textrm{leading}}:=\varepsilon_{\mu(s)} \, \alpha_{-1}^{\mu_1} \dots \alpha_{-1}^{\mu_s}\,,\quad T_{11} F_{\textrm{leading}} \ket{0;p} = 0 \,,
\end{align}
with
\begin{align}
    T^1{}_1 F_{\textrm{leading}} \ket{0;p} = s F_{\textrm{leading}} \ket{0;p} \,.
\end{align}
In other words, the spin $s$, namely the length of the row, is a \textit{free parameter} and proceeding along the trajectory amounts to adding one new box per level. Consequently, reaching arbitrarily high spin is possible \textit{without} exciting types of oscillators other than $\alpha_{-1}$.

To proceed to more complicated trajectories and given that the notion of a subleading trajectory had not previously been well--defined in the literature, two useful definitions, that facilitate the organization of the spectrum in terms of sets of infinite states in an efficient way, were made in \cite{Markou:2023ffh}. Firstly, because every (allowed) Young diagram will appear at its respective $N_{\textrm{min}}$, embedded within a Weinberg state, but also infinitely more times, albeit at higher mass levels $N$, namely \textit{deeper} in the spectrum (up to multiplicity), the notion of the \textit{depth} of a given Young diagram was defined as
\begin{align}
    w:= N- N_{\text{min}}\,.
\end{align}
So the depth $w$ is an integer parametrizing how deep inside the spectrum a certain Young diagram finds itself, hence also the \textit{complexity} of the respective polynomial w.\ r.\ t.\ the corresponding Weinberg state. All Weinberg states are thus depth $w=0$ states by definition. Considering then that every new row amounts to exciting a new oscillator at $w=0$, the notion of a \textit{trajectory} was defined as the set of diagrams with a fixed number of rows and at fixed depth $w$, up to multiplicity. The leading Regge trajectory is thus the set of all $1$--row diagrams at $w=0$, while the Young diagrams of the lightest member--states of the $2$--row trajectory at $w=0$ are denoted by a green outline in table \ref{table:two}. This $2$--row $w=0$ trajectory, whose polynomial is given by \eqref{simplest2}, has namely \textit{infinitely many branches}, whose number is parametrized by the length $s_2$ of the second row. For example, the first is the set of infinitely many diagrams with one box in the second row, starting with the antisymmetric tensor at level $N=3$ and proceeding along the branch amounts to adding boxes only to the \textit{first} row, with polynomial
\begin{align} \label{simplest2branch1}
 \Yboxdim{10pt}  \gyoung(_4{s_1},_1{1})\,,\quad   F_{\textrm{simpl}} = \varepsilon_{\mu(s_1),\, \lambda} \, \alpha_{-1}^{\mu_1} \dots \alpha_{-1}^{\mu_{s_1}} \,\alpha_{-2}^{\lambda}   \,.
\end{align}
The second branch is the set of infinitely many diagrams with two boxes in the second row, starting with the ``window'' at level $N=6$, while proceeding along the branch again amounts to adding boxes only to the \textit{first} row, with polynomial
\begin{align} \label{simplest2branch2}
 \Yboxdim{10pt}  \gyoung(_4{s_1},_2{2})\,,\quad   F_{\textrm{simpl}} = \varepsilon_{\mu(s_1),\, \lambda (2)} \, \alpha_{-1}^{\mu_1} \dots \alpha_{-1}^{\mu_{s_1}} \,\alpha_{-2}^{\lambda_1}  \alpha_{-2}^{\lambda_2}   \,.
\end{align}
The $3$--row $w=0$ trajectory starts with a branch which begins with the rank--$3$ antisymmetric tensor at level $6$ (denoted by an light blue outline), and so on. 
\begin{table}
\centering 
\renewcommand{\arraystretch}{1.5}
  \begin{tabular}{ c || l  }
   $N$ & decomposition in physical states  \\ \hline \hline
   $0$ & $\textcolor{BrickRed}{\bullet}$ \\
   $1$ & $\LeadingB{;}_{\textcolor{BrickRed}{so(D-2)}}$ \\
   $2$ & $ \LeadingB{;;}$ \\
   $3$ & $\LeadingB{;;;} \oplus \tworow{;,;}$ \\
   $4$ & $\LeadingB{;;;;} \oplus \tworow{;;,;} \oplus {\wtwo \gyoung(;;)}  \oplus \textcolor{Orange}{\bullet}$ \\
   $5$ & $\LeadingB{;;;;;} \oplus \tworow{;;;,;} \oplus {\wtwo  \gyoung(;;;)} \oplus {\wonesec \gyoung(;;,;)} \oplus {\wtwosec \gyoung(;,;)} \oplus {\wfour \gyoung(;)}$  \\
   $6$ & $\LeadingB{;;;;;;} \oplus \tworow{;;;;,;} \oplus {\wtwo \gyoung(;;;;)} \oplus {\wonesec \gyoung(;;;,;)} \oplus { \wtwosec \gyoung(;;,;)} \oplus {\wthree \gyoung(;;;)}  \oplus \tworow{;;,;;} \oplus \textcolor{Orange}{2}\, {\wfour \gyoung(;;)} \oplus \threerow{;,;,;} \oplus {\wfive \gyoung(;)} \oplus \bullet $
  \end{tabular}
\renewcommand{\arraystretch}{1}\caption{Colored outline and filling correspond to $w=0$ and to $w>0$ trajectories respectively, with different filling colors corresponding to different values of $w$ for a given number of rows.} \label{table:two}
\end{table}

Proceeding to $w>0$, we observe the lightest member--states of \textit{clones}, in terms of the respective Young diagrams, of the $w=0$ trajectories. For example, the trajectory of diagrams with  a red filling in table \ref{table:two} is a $1$--row $w=2$ trajectory, namely a clone of the leading Regge, albeit with a more complicated polynomial. Likewise the diagrams with a blue filling are the lightest member--states of the first branch of the $w=1$ clone of the previously discussed $2$--row $w=0$ trajectory, and so on. We already see how neatly an infinite number of branches is packaged together within a \textit{single} trajectory, for all trajectories with at least two rows, at any $w$. Moreover, we observe that there always appears a \textit{finite} number of trajectories at any value of the depth $w$, simply because an irreducible $\mathfrak{so}(D-1)$ Young diagram cannot have more than $D-1$ rows. This demonstrates how economical the aforementioned definitions of depth and trajectory are. The \textit{entire} open string spectrum thus splits in two parts: the Weinberg states, organized within $w=0$ trajectories with known polynomials, and all their ``clones'' at $w>0$, albeit with unknown polynomials. How can we systematically construct the latter?

Following  \cite{Markou:2023ffh}, let us add to the operators \eqref{eq:opT1} the operator
\begin{align}
    T^{k \ell} = \tfrac{1}{k \ell} \alpha_{-k} \cdot \alpha_{-\ell}\,.
\end{align}
By means of the oscillator algebra \eqref{eq:osc_alg}, it is straightforward to see that the set of operators $T^{k \ell}\,, T^k{}_\ell\,, T_{k \ell} $ satisfy the commutation relations
\begin{align} \label{sp1}
       [T^\ell{}_n,T^{km}]&= \delta^k_n T^{\ell m}+\delta^m_n T^{\ell k}\\ \label{sp2}
     [T_{km}, T^\ell{}_n]&= \delta^\ell_k T_{mn}+\delta^\ell_m T_{kn}\\ \label{sp3}
     [T^k{}_\ell,T^m{}_n]&= \delta_\ell^m T^k{}_n-\delta^k_n T^m{}_\ell\\ \label{sp4}
    [T_{km}, T^{\ell n}]&=(D-1)(\delta_k^n \delta_m^\ell + \delta_k^\ell \delta_m^n)+  \delta^\ell_k T^n{}_m+\delta^\ell_m T^n{}_k+\delta^n_k T^\ell{}_m+\delta^n_m T^\ell{}_k\,.
\end{align}
With $k,\ell,\ldots=1,2,\ldots,K$, namely $K$ being the range of the operators' indices, and upon the redefinition $T^k{}_\ell \rightarrow T^k{}_\ell + \tfrac{D-1}{2}  \delta^k_\ell\,$ in order to absorb the central term in \eqref{sp4}, it follows that the algebra \eqref{sp1}--\eqref{sp4} is (isomorphic to) the $\mathfrak{sp}(2K,\mathbb{R})$ algebra! This algebra operates on the \textit{labels}, namely the units of energy $n$ that the oscillators $\alpha_{-n}^\mu$ carry, and its generators are spacetime scalars. To cover the entire spectrum, all oscillators must be be considered, which essentially amounts to taking the \textit{inductive} or \textit{direct} limit
\begin{align}
   \mathfrak{sp} (2,\mathbb{R}) \subset \mathfrak{sp} (6,\mathbb{R}) \subset \mathfrak{sp} (8,\mathbb{R}) \subset \dots \,,\quad \varinjlim  \mathfrak{sp}(2K,\mathbb{R}) = \mathfrak{sp}(2 \bullet,\mathbb{R}) \,,
\end{align}
where the bullet symbol $\bullet$ now stands for infinity. But let us recall that all physical states' polarizations are irreps of the (little group of the) spacetime Lorentz algebra $\mathfrak{so}(D-1,1)$ that is generated by the Lorentz generators \eqref{eq:lorentzgen}. Since the $\mathfrak{so}$ and $\mathfrak{sp}$ algebras operate on different indices or, more formally for example (Exercise)
\begin{align}
   [ J^{\mu \nu}, T^k{}_\ell] = 0 \,,
\end{align}
we learn that they \textit{commute} or that they form what is known as a \textit{Howe dual pair}! In representation theory, Howe duality \cite{Howe1,Howe2}\footnote{See also \cite{Rowe:2012ym,Basile:2020gqi}.} relates, under specific conditions here satisfied, all irreps in the oscillator representation of two commuting algebras via a \textit{bijection}, so irreps form pairs in a ``$1$--$1$'' manner. In our context, this means that every $\mathfrak{sp}$ irrep is mapped to a unique Lorentz irrep and vice versa. 

The $\mathfrak{sp}$ generators may also be split in raising and lowering operators. A practical choice is to identify the raising (lowering) operators with those that also \textit{add} (\textit{remove}) units of energy, namely
\begin{align} 
\textrm{raising}&:\quad T^{k\ell}\,,\quad T^{k>\ell}{}_{\ell} \\
    \textrm{lowering}&:\quad T_{k\ell}\,,\quad T^{k<\ell}{}_{\ell} \,.
\end{align}
The operators $T^k{}_k$ then generate the Cartan subalgebra and the lowest weight states $F_{\textrm{l.w.}}\ket{0;p}$ of the $\mathfrak{sp}$ algebra can be defined via
\begin{align}
 T_{k\ell} F_{\textrm{l.w.}}\ket{0;p} =0\,,\quad T^{k<\ell}{}_{\ell} F_{\textrm{l.w.}}\ket{0;p} =0   \,,
\end{align}
together with the condition
\begin{align}
     T^{k}{}_{k} F_{\textrm{l.w.}}\ket{0;p} =  s_k  \,F_{\textrm{l.w.}}\ket{0;p} \,,
\end{align}
which fixes the $\mathfrak{sp}$ weight of the states $F_{\textrm{l.w.}}\ket{0;p}$, namely their eigenvalues w.\ r.\ t.\ the Cartan subalgebra generators (modulo the shift by $\tfrac{D-1}{2}$). We thus learn something fascinating, a key observation in \cite{Markou:2023ffh}: the Weinberg states, or the $w=0$ trajectories, \textit{are} the lowest weight states of the $\mathfrak{sp}$ algebra! Consequently, Howe duality may be applied in the following way: the infinitely many appearances of a given (and any) Young diagram correspond to a unique $\mathfrak{sp}$ irrep, the lowest weight state of which being mapped to the first appearance of the diagram, namely the one at the lowest possible level. Every other appearance of the diagram can thus be reached by moving inside the respective $\mathfrak{sp}$ irrep by means of the $\mathfrak{sp}$ raising operators. The $\mathfrak{sp}(2K,\mathbb{R})$ (unitary) irreps are infinite--dimensional, which is another way of saying that the multiplicity of every $\mathfrak{so}$ irrep in the Fock space is infinite! 

An efficient, algorithmic and covariant technology \cite{Markou:2023ffh} of constructing entire clones \textit{deeper} in the spectrum can thus be formulated as follows. To construct the polynomial  $F_{w>0}$ of a clone at a given depth $w$, one starts with the respective $w=0$ trajectory and dresses it by all possible combinations of the $\mathfrak{sp}$ raising operators that carry $w$ units of energy, as in
\begin{align} \label{ansatz}
    F_{w>0} =  f_{w>0}( T^{mn}, T^{k>\ell}{}_{\ell}) \, F_{w=0}\,.
\end{align}
Because of Howe duality, the dressing function $f_{w>0}$ does not alter the form of the Young diagrams of the $w=0$ trajectories, hence the cloning is possible. The Virasoro constrains must then be supplied with the Ansatz \eqref{ansatz} and solved to yield the clone including its multiplicity and here comes the enormous improvement in the efficiency of this method for the construction of the spectrum, compared to traditional techniques: by solving the Virasoro constraints \textit{only once}, infinitely many physical states are extracted at one go.

Let's consider a few examples in regard to the cloning of the leading Regge trajectory \eqref{leading}. At $w=1$, the Ansatz reads
\begin{align}
    f_{w=1}= \beta\, T^2{}_1 \,,
\end{align}
where $\beta$ is an a priori arbitrary parameter, so the $L_1$ constraint takes the form
\begin{align}
    \sum_{m=1} T^m{}_{1+m}  \, f_{w=1} \, F_{\textrm{leading}} \ket{0;p}= 0 \,,
\end{align}
which yield
\begin{align}
    \beta \, \big(T^1{}_2 + \ldots \big) \, T^2{}_1 \varepsilon_{\mu(s)} \alpha_{-1}^{\mu_1} \dots \alpha_{-1}^{\mu_s} \ket{0;p} = 0 \quad \Rightarrow \quad \beta \, s = 0\,,
\end{align}
where the dots do not contribute since they involve $\mathfrak{sp}$ lowering operators that commute with $T^2{}_1$ and annihilate the leading Regge and we have used the commutator \eqref{sp3} (the $L_2$ constraint yields no other restrictions). So it must be that $\beta = 0$, namely there exists no such trajectory\footnote{This result had actually been known for a long time, albeit shown without the use of the symplectic algebra and the technology, as pointed out to us by Paolo Di Vecchia.}! At $w=2$, the Ansatz reads,
\begin{align} \label{eq:ansatz2}
    f_{w=2}= \beta_1 \, T^{11} +\beta_2  T^3{}_1 + \beta_3 (T^2{}_1)^2 \,,
\end{align}
so the $L_1$ and $L_2$ constraints take respectively the form
\begin{align}
\big(T^1{}_2 + 2 T^2{}_3 + \ldots \big) \, f_{w=2} \, F_{\textrm{leading}} \ket{0;p} = 0 \,,\quad       \big(T_{11} + 2 T^1{}_3 + \ldots \big)  \, f_{w=2} \, F_{\textrm{leading}} \ket{0;p}  = 0\,,
\end{align}
which yields
\begin{align} \label{eq:solw=2lead}
    \beta_2= -\beta_1 \frac{D+2s-1}{s} \,,\quad \beta_3 = \beta_1 \frac{D+2s-1}{s(s-1)} \,,
\end{align}
where the spin $s$ is a free parameter, namely we have extracted the \textit{entire} clone at $w=2$. Because $\beta_1$ has become an overall prefactor, there exists only one clone of the leading Regge at this depth, while the divergences in the parametrization \eqref{eq:solw=2lead} of the solution indicate what the lightest member--state of the clone is: the states with $s=0,1$ are not allowed, in accordance with what we observe in table \ref{table:two}. Indeed, for $s=0$, the Ansatz \eqref{eq:ansatz2} only contains the first term, and the Virasoro constraints impose $\beta_1=0 $, while for $s=1$ only the first two terms of the Ansatz \eqref{eq:ansatz2} contribute, with the Virasoro constraints imposing $\beta_1 = 0 = \beta_2$. To showcase the technology's efficiency, let us also substitute the solution \eqref{eq:solw=2lead} in the Ansatz \eqref{eq:ansatz2} to write explicitly the form of the entire physical clone, namely
\begin{align} \label{eq:clone2}
\begin{aligned}
    F_{w=2}  &= \varepsilon_{\mu(s)} \Big[\alpha_{-1} \cdot \alpha_{-1} \alpha_{-1}^{\mu_1} \dots \alpha_{-1}^{\mu_s} -\tfrac{25+2s}{3} \, \alpha_{-1}^{\mu_1} \dots \alpha_{-1}^{\mu_{s-1}} \alpha_{-3}^{\mu_s}  \\
& \qquad  \qquad    +\tfrac{25+2s}{4} \, \alpha_{-1}^{\mu_1} \dots \alpha_{-1}^{\mu_{s-2}} \alpha_{-2}^{\mu_{s-1}}  \alpha_{-2}^{\mu_{s}}   \Big]   \,, \quad s \ge 2\,.
\end{aligned}
\end{align}
It is now trivial to compare the complexity of the polynomial for $s=2$ with its respective Weinberg state \eqref{eq:lightestspin2} (Exercise).

To conclude, the technology reviewed in this section is an algorithmic, covariant and efficient method for the construction of entire trajectories deeper in the open bosonic string spectrum. Once the trajectory and the depth to clone it at is chosen, extracting the clone(s) amounts to writing the Ansatz and solving the Virasoro constraints. For fixed depth $w$, the Ansatz always contains a \textit{finite} number of terms built out of the $\mathfrak{sp}$ raising operators, but different subalgebras of the larger $\mathfrak{sp}(2K,\mathbb{R})$ may be relevant, depending on the trajectory undergoing cloning. For example, the $w=2$ clone \eqref{eq:clone2} of the leading Regge involves exciting up to the $\alpha_{-3}^\mu$ oscillator, so the algebra generating the clone is $\mathfrak{sp}(6,\mathbb{R})$, while it is straightforward to see that it is $\mathfrak{sp}(8,\mathbb{R})$ that generates the clones of the $2$--row $w=0$ trajectory at $w=2$. It is further worth highlighting that the Virasoro constraints contain all physical information about the clones: they reveal whether the trajectory in question even has a clone at the depth under probe and, in that case, they yield its physical polynomial including the state at which it is truncated, namely what its lightest member--state is. The number of solutions for the parameters' of the Ansatz becomes then the \textit{multiplicity} of the clones, and indeed there are such examples for more complicated trajectories. The efficiency of the technology is due to the possibility of extracting infinitely many states at one go and, recalling that e.\ g.\ the $2$--row trajectory at $w=0$ contains infinitely many branches, embedding it at a depth $w>0$ amounts to extracting \textit{infinitely many sets of infinitely many states} at once. The mathematical robustness of the technology is also manifest, since Howe duality guarantees that the entire spectrum can be constructed in this way, as it relates \textit{irreducible} representations. Of course, it remains an open problem to solve the constraints for \textit{arbitrary} depth, but the technology appears to offer the tools to address this question.  Let us finally stress that the symplectic algebra is not a symmetry of the spectrum, as not all Ans\"atze are allowed by the Virasoro constraints, which is why we refer to it as a spectrum--generating algebra instead\footnote{We thank Vasilis Niarchos for a related clarification.}. 

%%%%%%%%%%%%%%%%%%%%%%%%%%%%%%%%%%%%%%%%%%%%%%%%%%%%%%%%%%%%%
\section{CFT basics and elements of string interactions}
\label{sec:int}
%%%%%%%%%%%%%%%%%%%%%%%%%%%%%%%%%%%%%%%%%%%%%%%%%%%%%%%%%%%%%

Until now we have used the ``global objects'' $\alpha_{-n}^\mu$ to construct string states. For string interactions, we will turn to the residual worldsheet symmetry, namely use the language of conformal field theory to define ``local operators'' that are the ingredients for the modern computation of string scattering amplitudes. Everything we have seen so far in terms of the construction of the perturbative spectrum has a realization in the CFT language \cite{Belavin:1984vu,Cardy:1989ir} (and \cite{Markou:2023ffh} contains the CFT version of the technology), of which we will briefly review the rudiments. To begin with, let us consider the \textit{closed string} worldsheet,  namely the cylinder parametrized by $\sigma$ and $\tau$. A Wick rotation allows to switch to Euclidean time on the worldsheet, so that now
\begin{align}
    \sigma^\pm = -\mathrm{i} (\tau\pm \mathrm{i} \sigma)\,,
\end{align}
after which we may define the complex cylinder via
\begin{align}
 w = \tau - \mathrm{i} \sigma \,,\quad \bar w = \tau + \mathrm{i} \sigma\,.
\end{align}
A conformal transformation can then be used to switch to the complex plane $SL(2,\mathbb{C})$
\begin{align}
    z= e^w \,,\quad \bar z = e^{\bar w}\,,
\end{align}
or, more accurately, to the Riemman sphere $\mathbb{C} \cup \{\infty\}$. Concentric circles on the complex plane correspond then to constant time $\tau$, with their center ($z \rightarrow 0$) corresponding to the infinite past ($\tau \rightarrow - \infty$) and the point at infinity ($z \rightarrow \infty$) corresponding to the infinite future ($\tau \rightarrow \infty$). The Polyakov action then takes the form
\begin{align} \label{eq:polcomplex}
    S_{\textrm{P}}^{\textrm{c.g.}} = \frac{1}{\pi \alpha'} \int \md^2z \, \partial X(z,\bar z) \cdot \bar \partial X (z,\bar z)\,, \quad \partial := \partial_z \,,\quad \bar \partial := \partial_{\bar z}
\end{align}
and the e.\ o.\ m.\ of the field $X^\mu$ takes the form 
\begin{align}
    \partial \bar{\partial} X (z,\bar z) = 0 \quad \Rightarrow \quad  X(z,\bar z) = X(z) +\bar{X}(\bar z), 
\end{align}
so we may expand the solution in Fourier modes as (Exercise)
\begin{align}
    X^\mu (z,\bar z) = x^\mu -\tfrac{\mathrm{i}}{2} \alpha' p^\mu \ln |z|^2 + \mathrm{i} \sqrt{\tfrac{\alpha'}{2}} \sum_{n \neq 0} \frac{1}{n} \bigg(\frac{\alpha_n^\mu}{z^n} + \frac{\tilde{\alpha}_n^\mu}{\bar{z}^n} \bigg)\,.
\end{align}

Let us consider the derivatives
\begin{align} \label{eq:primary}
 \partial X^\mu (z,\bar z) =  \partial X^\mu (z) = -\mathrm{i} \sqrt{\tfrac{\alpha'}{2}} \sum_{n\in \mathbb{Z}} \frac{\alpha_n^\mu}{z^{n+1}} \,, \quad   \bar  \partial X^\mu (z,\bar z) =     \bar \partial \bar{X}^\mu (\bar z) = -\mathrm{i} \sqrt{\tfrac{\alpha'}{2}} \sum_{n\in \mathbb{Z}} \frac{\tilde{\alpha}_n^\mu}{\bar{z}^{n+1}} \,.
\end{align}
The $2$--point correlation function on the sphere $S^2$ may then be defined as
\begin{align}
    \langle \partial_z X^\mu (z) \partial_w X^\nu (w) \rangle_{S^2} =  -\tfrac{\alpha'}{2} \sum_{n,m \in \mathbb{Z}} z^{-n-1} w^{-m-1} \bra{0;p} \alpha_n^\mu \alpha_m^\nu \ket{0;p} \,,
\end{align}
where we postpone considering the ordering of the operators in the LHS for a moment. In the RHS, only the $\alpha_{-m}^\nu$ and $\alpha_{n}^\mu$ for $n,m>0$ contribute, so we have
\begin{align}
\sum_{n,m \in \mathbb{Z}} z^{-n-1} w^{-m-1} \bra{0;p} \alpha_n^\mu \alpha_m^\nu \ket{0;p} =    \eta^{\mu \nu} \frac{1}{wz} \sum_{n=0}^\infty n\, \Big(\frac{w}{z} \Big)^n\,.
\end{align}
With a simple trick (Exercise), we can bring the sum to the form of the geometric series
\begin{align}
    \sum_{n=0}^\infty r^n = \frac{1}{1-r}\,, \quad r = w/z\,,
\end{align}
which converges only for $|r|<1$! Consequently, for products $ A(z) A(w) $ of a field $A(z)$ on the complex plane to be well--defined, we must impose the \textit{radial ordering}
\begin{align}
    |z|>|w| \,,
\end{align}
which is the analogue of \textit{time ordering} on the cylinder. We then have that
\begin{align} \label{eq:2pointsphere}
        \langle \partial_z X^\mu (z) \partial_w X^\nu (w) \rangle_{S^2} = - \tfrac{\alpha'}{2} \eta^{\mu \nu} \frac{1}{(z-w)^2}
\end{align}
and an analogous expression for the correlator of the field $\bar{\partial} \bar{X}^\mu(\bar z)$. Importantly, the result \eqref{eq:2pointsphere} can also be derived from the propagator
\begin{align} \label{eq:sphereprop}
    \langle X^\mu(z, \bar z) X^\nu(w,\bar w) \rangle_{S^2}  = - \tfrac{\alpha'}{2} \eta^{\mu \nu}  \ln |z-w|^2 \,,
\end{align}
which can be deduced from the action \eqref{eq:polcomplex} (Exercise). These results imply that we will be able to calculate correlation functions on the worldsheet.

Moreover, the expressions \eqref{eq:primary} imply that the oscillators $\alpha_n^\mu$ and $\tilde{\alpha}_n^\mu$ can be thought of as the \textit{Laurent coefficients} of the fields $\partial X^\mu$ and $\bar \partial X^\mu$ respectively, so using the Cauchy integral formula we may invert the expressions \eqref{eq:primary} to find
\begin{align} \label{eq:Laurentcoeff}
    \alpha_n^\mu = \mathrm{i} \sqrt{\tfrac{2}{\alpha'}} \oint \frac{\md z}{2\pi \mathrm{i}} \partial X^\mu (z) \,z^n
\end{align}
and similarly for $\tilde{\alpha}_n^\mu$.
Restricting to the creation operators, we then have that 
\begin{align} \label{eq:diction}
    \alpha_{-n}^\mu \ket{0} = \mathrm{i} \sqrt{\tfrac{2}{\alpha'}}  \oint \frac{\md z}{2 \pi \mathrm{i}} \frac{\partial X^\mu(z)}{z^n} \ket{0} = \mathrm{i} \sqrt{\tfrac{2}{\alpha'}} \frac{1}{(n-1)!} \partial^n X^\mu (0) \ket{0}\,, \quad n>0\,,
\end{align}
where we have used the Cauchy differentiation formula. The relation \eqref{eq:diction} provides a \textit{dictionary} between a creation operator carrying $n$ units of energy and the $n$--th derivative of the field $X^\mu$ evaluated at $z=0$, namely in the \textit{infinite past}. This is a hint that we will be able to prepare \textit{asymptotic} string states as worldsheet fields in order to compute scattering amplitudes!

Another insightful computation is to consider the analogue of the oscillator algebra \eqref{eq:osc_alg} in the CFT language. Using the expressions \eqref{eq:Laurentcoeff}, we have that
\begin{align}
    [\alpha_m^\mu,\alpha_n^\nu] = -\tfrac{2}{\alpha'} \oint_{C_0^z} \frac{\md z}{2\pi \mathrm{i} } \oint_{C_0^w} \frac{\md w}{2\pi \mathrm{i}}   \, \Big(\partial X^\mu(z) \partial X^\nu (w) - \partial X^\nu(w) \partial X^\mu(z) \Big) \, z^m w^n\,,
\end{align}
where the symbol $C_0^z $ stands for a circle parametrized by $z$ with its center at the point $(0,0)$ of the complex plane. In the RHS, if we focus on the integration w.\ r.\ t.\ the coordinate $z$, radial ordering forces $|z|>|w|$ for the first term and $|w|>|z|$ for the second term, so the subtraction of latter from the former amounts to an integral around a circle parametrized by $z$ and centered at $w$, namely 
\begin{align}
    [\alpha_m^\mu,\alpha_n^\nu] = -\tfrac{2}{\alpha'}  \oint_{C_0^w} \frac{\md w}{2\pi \mathrm{i} }  \oint_{C_w^z} \frac{\md z}{2\pi \mathrm{i} }  \, \partial X^\mu(z) \partial X^\nu (w) \, z^m w^n\,.
\end{align}
This means that we have to consider the product $\partial X^\mu(z) \partial X^\nu (w)$ more carefully. In CFT language, such an object can be written as an Operator Product Expansion (OPE), namely it can be expanded in local operators. The OPE is always a sum of two terms (each of which may be a sum of terms): a \textit{regular} part, that is given by the normal--ordered product, and a \textit{singular} part, which contains poles and for the example of $\partial X^\mu(z) \partial X^\nu (w)$ is accounted for by the $2$--point function \eqref{eq:2pointsphere}, namely
\begin{align} \label{eq:ope}
   \partial X^\mu(z) \partial X^\nu (w) =  - \tfrac{\alpha'}{2} \eta^{\mu \nu} \frac{1}{(z-w)^2} + \textrm{regular terms}\,.
\end{align}
Expanding $z$ around $w$, we thus have
\begin{align} \label{eq:doubleint}
\begin{aligned}
 \relax       [\alpha_m^\mu,\alpha_n^\nu] & = -\tfrac{2}{\alpha'}  \oint_{C_0^w} \frac{\md w}{2\pi \mathrm{i} }\, w^n \oint_{C_w^z} \frac{\md z}{2\pi \mathrm{i} }  \, \Big(  - \tfrac{\alpha'}{2} \frac{\eta^{\mu \nu} }{(z-w)^2}   + \textrm{regular terms}\Big)  \\
& \hspace{6cm} \times       \big(w^m + m (z-w)w^{m-1} + \ldots\big) \,.
\end{aligned}
\end{align}
Since the regular terms and the dots do not contribute to the $z$ integral, there is only a contribution from a second-- and a first--order pole and, using again the Cauchy differentiation formula it is easy to see that the value of the double integral in the RHS of \eqref{eq:doubleint} becomes the RHS of the commutator \eqref{eq:osc_alg}! Consequently, we may think of the analogue of the oscillator algebra as being the OPE \eqref{eq:ope}.

Similarly, we may wonder what the analogue of the Virasoro algebra \eqref{eq:Vir_alg} is. Let us consider the worldsheet energy--momentum tensor, which on the complex plane now reads
\begin{align}
 T(z) := - \tfrac{1}{\alpha'} :\partial X \cdot \partial X: \,,\quad  \bar{T}(\bar z) = -\tfrac{1}{\alpha'} :\bar{\partial} X \cdot \bar{\partial X}:\,.
\end{align}
It can then be shown (Exercise) that the Virasoro modes $L_n$ and $\tilde{L}_n$ are the Laurent coefficients of the energy--momentum tensor, namely
\begin{align}
    T(z)  = \sum_{n\in \mathbb{Z}} L_n \frac{1}{z^{n+2}}
\end{align}
and similarly for $\bar{T}(\bar z)$. The analogue of the Virasoro algebra is then the OPE (Exercise)
\begin{align} \label{eq:opevir}
    T(z) T(w) \sim \frac{D/2}{(z-w)^4} +\frac{2T(w)}{(z-w)^2} + \frac{\partial T(w)}{z-w}\,,
\end{align}
where the similarity symbol means that only the singular part of the OPE is displayed in the RHS.

The form of the aforementioned OPEs in fact encodes the information as to how the fields involved \textit{transform} under conformal transformations. More specifically, a particular class of fields of a $2$--dimensional CFT is that of \textit{conformal} or \textit{primary} fields $\phi (z,\bar z)$, which under conformal transformations
\begin{align}
    z \rightarrow z'= f(z) \,,\quad \bar z \rightarrow \bar{z}' = \bar{f}(\bar z)
\end{align}
transform as
\begin{align}
    \phi (z,\bar z) \rightarrow \phi'(z',\bar z') = \bigg( \frac{\partial z'}{\partial z}\bigg)^{-h} \bigg( \frac{\partial \bar z'}{\partial \bar z}\bigg)^{- \bar h} \phi(z,\bar z)\,,
\end{align}
where $h$ and $\bar h$ are the \textit{conformal weights} of the field $\phi(z,\bar z)$ (and are not necessarily conjugate to each other). This property is equivalent to the field $\phi(z)$ having an expansion in the complex plane of the form
\begin{align}
    \phi (z) = \sum_{n \in \mathbb{Z}} \frac{\phi_n}{z^{n+h}}
\end{align}
and similarly for $\bar{\phi}(\bar z)$, if  $\phi(z,\bar z)= \phi(z) + \bar{\phi}(\bar z)$, where $\phi_n$ and $\tilde{\phi}_n$ are the Laurent coefficients. Moreover, under infinitesimal conformal transformations
\begin{align}
    z\rightarrow z + \xi(z)\,,
\end{align}
the variation of the field $\phi(z)$ reads
\begin{align} \label{eq:variation}
    \delta \phi(z) = - (h \partial \xi + \xi \partial) \phi \overset{!}{=}      - \big[  T_\xi, \phi(z) \big]\,,
\end{align}
since $T(z)$ is the generator of conformal transformations, with $T_\xi$ being the respective charge, namely
\begin{align}
    T_\xi = \oint_{C^w_0}\xi(w) T(w)\,.
\end{align}
Evaluating the commutator in \eqref{eq:variation} as before, we find that the equality holds iff the OPE of the field $\phi(z)$ with the energy--momentum tensor takes the form
\begin{align}
    T(z) \phi(w) \sim  \frac{h\, \phi(w)}{(z-w)^2} + \frac{\partial \phi(w)}{z-w} \,,
\end{align}
which can be seen as alternative definition for a primary field of weight $h$. We thus learn that $\partial X^\mu$ is a primary field of weight $1$ (unlike the field $X^\mu(z)$, as can be shown), while its \textit{descendants} $\partial^n X^\mu$ have weight $n$; the conformal weight is the analogue of the units of energy that the creation oscillators carry due to the dictionary \eqref{eq:diction}. Moreover, we learn that the energy--momentum tensor is \textit{not} a primary field, given the presence of the $4$--th order pole in the OPE \eqref{eq:opevir}. 

So far we have focused on the closed string CFT, in the context of which the field $X^\mu$ split in a holomorphic and an antiholomorphic part, with all OPEs and 2--point functions discussed having a holomorphic and an antiholomorphic version. The open string, however, has a boundary and the relevant CFT is a \textit{boundary} CFT, with the relevant worldsheet topology becoming essentially that of the disk $\mathbb{D}_2$ with the real line $\mathbb{R}$ as its boundary, or eqiuvallently $SL(2,\mathbb{R})$, namely the upper half plane. Open string correlators can then be obtained from the closed string ones by means of Cardy's trick, known also as the \textit{doubling trick}, that was previously referred to in the oscillator language as in \eqref{eq:doubling}, according to which
\begin{align} \label{eq:diskprop}
    \langle X^\mu(z) X^\nu(w) \rangle_{\mathbb{D}_2}  = - 2\alpha' \eta^{\mu \nu}  \ln |z-w| \,,
\end{align}
whose proof we do not reproduce here. For our purposes, and if we restrict to open string amplitudes only (and for spacetime--filling D--branes), the doubling trick also simply amounts to making the replacement $\alpha' \rightarrow 4 \alpha' $ in the boson $2$--point function \eqref{eq:2pointsphere}, namely 
\begin{align} \label{eq:2pointdisk}
        \langle \partial_z X^\mu (z) \partial_w X^\nu (w) \rangle_{\mathbb{D}_2}= - 2\alpha' \eta^{\mu \nu} \frac{1}{(z-w)^2}
\end{align}
(with no antiholomorphic counterpart). Open string asymptotic states can then be defined by means of the dictionary \eqref{eq:diction} via the \textit{state--operator correspondence}, namely
\begin{align}
\ket{\textrm{phys}} = \lim_{z \rightarrow 0 } V(z) \ket{0}
\end{align}
where $ \ket{\textrm{phys}}$ is an asymptotic state built out of the creation operators $\alpha_{-1}^\mu\,,\alpha_{-2}^\nu\,,\ldots$ as previously discussed and $V(z)$ is a \textit{primary} field that is a function $F$ of the primary field $\partial X$ and its descendants containing also the momentum eigenstate $e^{\mathrm{i} p \cdot X(z)}$. The field $V(z)$ is a spacetime scalar (formed by contracting the spacetime indices of $\partial^n X^\mu$ with the state's polarization, like in the oscillator language) and is called the \textit{vertex operator} that creates a string state at a point $z$ of the worldsheet.

The analogue of the Virasoro constraints \eqref{eq:Vir_suff} becomes the condition
\begin{align} \label{eq:physCFT}
    [Q,V(z)] = \textrm{tot.\ deriv.\ } \,,
\end{align}
where $Q$ is the worldsheet nilpotent BRST charge, a notion that the student reader may be familiar with from a QFT course. Here, it is defined as
\begin{align}
    Q := \oint \frac{\md z}{2\pi \mathrm{i}} c(z) \Big[ T(z) + \big(\partial c(z)\big) b(z) \Big] \,,
\end{align}
where $(b,c)$ is a system of anticommuting\textit{ Faddeev--Popov ghosts} with $3$--point function
\begin{align} \label{eq:ghost3}
    \langle  c(z) c(y) c(w) \rangle_{\mathbb{D}_2} = |(z-y)(z-w)(y-w)|\,,
\end{align}
which is introduced in the BRST quantization of the string to \textit{cancel the Weyl anomaly}. Indeed, the central charge $c_{\textrm{ghost}}$ of the $(b,c)$ system turns out to be equal to $-26$, canceling the central charge of the CFT of the field $\partial X^\mu$. The physicality condition \eqref{eq:physCFT} firstly implies that all physical vertex operators have conformal weight $1$ (Exercise). Since the momentum eigenstate $e^{\mathrm{i} p\cdot X(z)}$ has conformal weight $\alpha' p^2=1-N$ (Exercise), the analogue of constructing string states on a level--by--level basis becomes writing an Ansatz for the function $F$ with all possible combinations of $\partial^n X$ that carry $N$ units of conformal weight and then solving the remaining conditions implied by \eqref{eq:physCFT}. More simply, all states discussed in the previous section can be translated into the respective vertex operators by means of the dictionary \eqref{eq:diction}. For example, the tachyon's vertex operator takes the simple form
\begin{align} \label{eq:vectorvo}
    V(z,p) =\tfrac{1}{\sqrt{2\alpha'}} g_{\textrm{o}} T^a \, e^{ip \cdot X(z)}  \,,\quad p^2 = \tfrac{1}{\alpha'}\,,
\end{align}
while the massless vector's \eqref{eq:masslessv} vertex operator reads
\begin{align} \label{eq:vectorvo}
    V^{a}(z,p) =\tfrac{1}{\sqrt{2\alpha'}} g_{\textrm{o}} T^a \, \varepsilon \cdot \partial X(z)  \, e^{ip \cdot X(z)} \,,\quad \varepsilon \cdot p =0 \,,\quad p^2 = 0\,.
\end{align}

Computing tree--level open string amplitudes amounts then to calculating the \textit{correlation function} of (all possible orderings of) the vertex operators of the external legs and integrating the result over all possible positions of the vertex operators insertions' on the string worldsheet, as they are unphysical. Because of conformal symmetry, the measure of the integral contains a division by the volume  $V_{\textrm{CKG}}$ of the conformal Killing group, namely $SL(2,\mathbb{R})$ for open strings. Its contribution is treated by fixing the insertion location of \textit{three} vertex operators and inserting  \textit{three} $c$--ghosts in the respective locations in the correlation function, with the explanation being outside the scope of these lectures. Let us consider the example of the scattering of three massless vectors \eqref{eq:vectorvo}.
The amplitude can be written as
\begin{align} \label{eq:3point}
\begin{aligned}
  \mathcal{A}_3 & = \int \frac{\md z_1 \md z_2 \md z_3}{V_{\textrm{CKG}}} \langle \, :c(z_1)\,V^{a_1}(z_1,p_1):\,: c(z_2) \, V^{a_2}(z_2,p_2):\,: c(z_3) \,V^{a_3}(z_3,p_3) :  \, \rangle_{\mathbb{D}_2} \\
 & \hspace{3cm}   + 2 \leftrightarrow 3\,,
\end{aligned}
\end{align}
namely it's a sum of the two possible orderings of three vertex operators, with the three $c$ ghosts explicitly inserted. The CFTs of the $(b,c)$--ghost system and of the field $\partial X^\mu$ are uncoupled, so to compute the correlator in \eqref{eq:3point} we can use the ghost $3$--point function \eqref{eq:ghost3} as well as compute the correlator
\begin{align} \label{eq:correthree}
\mathcal{A}^{\mu \nu \lambda} := \langle \, : \partial X_1^\mu  \, e^{\mathrm{i} p_1 \cdot X_1} :\,:  \partial X_2^\nu  \, e^{\mathrm{i} p_2 \cdot X_2} :\,: \partial X_3^\lambda  \, e^{\mathrm{i}p_3 \cdot X_3}: \,  \rangle_{\mathbb{D}_2}\,,
\end{align}
where we use the notation $X_i := X(z_i)$. We then have (on the disk)
\begin{align}
\begin{aligned}
  \mathcal{A}^{\mu \nu \lambda} &=  \sum_{n,k,\ell} \tfrac{1}{n!k!\ell!} \langle \, :  \partial X_1^\mu  \, \big(\mathrm{i} p_1 \cdot X_1\big)^n :\,:  \partial X_2^\nu  \, \big(\mathrm{i} p_2 \cdot X_2\big)^k :\,:  \partial X_3^\lambda  \, \big(\mathrm{i} p_3 \cdot X_3 \big)^\ell : \,  \rangle \\
  &  = \mathcal{E}_3 \Big[ \langle \mathrm{i}  \partial X_1^\mu \mathrm{i}  \partial X_2^\nu \rangle \Big(\langle \mathrm{i}  p_1 \cdot X_1 \, \mathrm{i}  \partial X_3^\lambda \rangle + \langle \mathrm{i}  p_2 \cdot X_2 \, \mathrm{i}  \partial X_3^\lambda \rangle \Big) \\
   & \quad \qquad + \langle \mathrm{i}  \partial X_1^\mu \mathrm{i}  \partial X_3^\lambda \rangle \Big(\langle \mathrm{i}  p_1 \cdot X_1 \, \mathrm{i}  \partial X_2^\nu \rangle + \langle \mathrm{i}  \partial X_2^\nu \, \mathrm{i}  p_3 \cdot X_3  \rangle \Big)   \\
   & \quad \qquad  + \langle \mathrm{i}  \partial X_2^\nu \mathrm{i}  \partial X_3^\lambda \rangle \Big(\langle \mathrm{i}  \partial X_1^\mu \, \mathrm{i}  p_2 \cdot X_2  \rangle + \langle \mathrm{i}  \partial X_1^\mu \,  \mathrm{i}  p_3 \cdot X_3  \rangle \Big) + 3\textrm{-momenta terms} \Big] \,,
\end{aligned}
\end{align}
where we have used Wick's theorem to perform the contractions, with the $3$--momenta terms arising from contracting all $\partial X$'s with an exponential. We have also set
\begin{align}\label{eq:kb3}
\mathcal{E}_3 := \langle \, :  e^{\mathrm{i} p_1 \cdot X_1} :\,:   e^{\mathrm{i} p_2 \cdot X_2} :\,:  e^{\mathrm{i}  p_3 \cdot X_3}: \,  \rangle_{\mathbb{D}_2} = \prod_{i<j}^3 |z_{ij}|^{2\alpha' p_i \cdot p_j} 
\end{align}
and the proof of this equality we will discuss in the next section.

We can then use momentum conservation
\begin{align}
    p_1^\mu + p_2^\mu +p_3^\mu =0
\end{align}
as well as the transversality and spins of the external legs to show that $\mathcal{E}_3=1$ for the three massless vectors. Substituting then the propagator \eqref{eq:diskprop}, the $z$--dependence of the amplitude cancels out completely and there is no integration left to perform, which is in fact a \textit{generic property} of 3--point string amplitudes regardless of external states.  Up to an overall prefactor that depends on the string coupling, the amplitude then takes the following form \textit{to lowest order} in momenta (or $\alpha'$),
\begin{align}
\begin{aligned}
     \mathcal{A}_3 & = \big( \Tr(T^{a_1} T^{a_2} T^{a_3}) \, \varepsilon_1 \cdot \varepsilon_2 \,  p_2 \cdot \varepsilon_3  + \textrm{cyclic permutations} \big) + 2\leftrightarrow 3 \,,
\end{aligned}
\end{align}
where the trace of the three Chan--Paton factors appears because the endpoints of the three strings are identified in pairs during the scattering process, so that the three states interact. Using again momentum conservation and transversality, we then have
\begin{align}
\begin{aligned}
     \mathcal{A}_3      &  = \Big[ \varepsilon_1 \cdot \varepsilon_3 \Big( \Tr(T^{a_1} T^{a_2} T^{a_3})  \,  p_1 \cdot \varepsilon_2 +  \Tr(T^{a_1} T^{a_3} T^{a_2})  \,  p_3 \cdot \varepsilon_2   \Big) \Big] + \textrm{cyclic} \\ \label{eq:YM3}
     & \sim f^{a_1 a_2 a_3} \, \varepsilon_1 \cdot \varepsilon_3 \, p_3 \cdot \varepsilon_2 + \textrm{cyclic} \,,
\end{aligned}
\end{align}
which is precisely the $3$--point amplitude in Yang--Mills theory with structure constants $ f^{a_1 a_2 a_3}$! More generally, the low--energy interactions of the open string's massless vector are of the YM type as shown by Neveu and Scherk \cite{Neveu:1971mu}, with the string corrections depending on higher orders in momenta (or $\alpha'$).

Interestingly, by setting
\begin{align}
    \varepsilon_{\mu \nu} := 2 \varepsilon_{(\mu} \tilde{\varepsilon}_{\nu)} \,,\quad \varepsilon \cdot \tilde \varepsilon = 0
\end{align}
and taking the \textit{product} of two amplitudes \eqref{eq:YM3},
\begin{align}
\begin{aligned}
 \mathcal{A}_3 \tilde{\mathcal{A}}_3 & \sim \big(\varepsilon_1 \cdot \varepsilon_3 \, p_3 \cdot \varepsilon_2 + \textrm{cyclic}  \big) \big(\tilde{\varepsilon}_1 \cdot \tilde{\varepsilon}_3 \, p_3 \cdot \tilde{\varepsilon}_2 + \textrm{cyclic}  \big) \\
 & \sim \Tr ( \varepsilon^1 \cdot \varepsilon^3 )\,  p_3 \cdot \varepsilon_2 \cdot p_3 + 2 p_1 \cdot\varepsilon_3 \cdot \varepsilon_1 \cdot \varepsilon_2 \cdot p_3 + \textrm{cyclic}  \sim \mathcal{M}_3\,,
\end{aligned}
\end{align}
where $\Tr ( \varepsilon^1 \cdot \varepsilon^3 ) = \varepsilon^1_{\mu \nu} \varepsilon^{3\, \mu \nu}$, the $3$--\textit{graviton} amplitude $\mathcal{M}_3$ emerges! In the above, we have identified the string endpoints with each other for each of the three strings, such that three closed strings that are not charged under the brane gauge group (since $\Tr(T^a T^b)$ is just a number) are formed. Each of the three closed strings is then associated with a closed string state formed as the \textit{tensor product} of two vectors, which contains the graviton in its symmetric traceless component. Its vertex operator then reads
\begin{align}
    V(z,\bar z,p) = \varepsilon_{\mu \nu} \partial X^\mu(z) \bar{\partial}  \bar{X}^\nu(\bar z) e^{\mathrm{i}p\cdot X(z,\bar z)}\,, \quad p^\mu \varepsilon_{\mu \nu} = 0 = \varepsilon^\mu{}_\mu \,,\quad p^2 =0\,.
\end{align}
More generally, the lowest--order interactions of the closed string's massless (symmetric traceless) rank--$2$ tensor match precisely those of the graviton of General Relativity predicted by the Einstein--Hilbert action \cite{Yoneya:1973ca,Yoneya:1974jg,Scherk:1974ca} and this imposes a relation between the Planck and the string scale $\alpha'$, by comparing the prefactors of field and string amplitudes to lowest order in $\alpha'$. In fact, this is the simplest (almost trivial) example of the powerful \textit{double copy} structure of Bern, Carrasco and Johansson \cite{Bern:2008qj,Bern:2010ue} that relates gauge theory with gravitational amplitudes, as well as of its stringy origin away from the $\alpha' \rightarrow 0$ limit, the Kawai, Lewellen and Tye relations between open and closed string amplitudes \cite{Kawai:1985xq}.

%%%%%%%%%%%%%%%%%%%%%%%%%%%%%%%%%%%%%%%%%%%%%%%%%%%%%%%%%%%%%
\section{The generating functional and scattering subleading trajectories*}
\label{sec:genf}
%%%%%%%%%%%%%%%%%%%%%%%%%%%%%%%%%%%%%%%%%%%%%%%%%%%%%%%%%%%%%

The correlator $\mathcal{E}$, namely the correlation function of the momentum eigenstates of the external legs, that is essentially the \textit{tachyon amplitude}, also known as the Koba--Nielsen factor, is an omnipresent factor in tree--level string scattering amplitudes \cite{Koba:1969rw,Koba:1969kh}. A way to compute it is by means of the generating functional method, which we know outline for the case of three tachyons, following in part \cite{Sagnotti:2010at}. By employing the string path integral, we may write the Koba--Nielsen factor as
\begin{align}
\begin{aligned}
    \mathcal{E}_3 & := \langle \, :  e^{\mathrm{i} p_1 \cdot X_1} :\,:   e^{\mathrm{i} p_2 \cdot X_2} :\,:  e^{\mathrm{i}p_3 \cdot X_3}: \,  \rangle_{\mathbb{D}_2} \\
 & \sim \int \mathcal{D} [X] e^{\mathrm{i} S_{\textrm{P}}^{\textrm{c.g.}} [X]} \prod_{i=1}^3  :e^{\mathrm{i} p_i \cdot X_i}: \\
& =  \int \mathcal{D}[X] \exp \bigg[ - \int \md^2z \bigg(  \tfrac{\mathrm{i}}{\pi \alpha'} \,  X^\mu(z)  \partial \bar \partial \bar{X}_\mu (\bar z) - \mathrm{i} \sum_{i=1}^3 p_i^\mu \delta^2(z-z_i) X_\mu(z) \bigg)  \bigg]\,,
\end{aligned}
\end{align}
where we can recognize the \textit{current}
\begin{align} \label{eq:current}
    J^\mu(z) = \sum_{i=1}^3 p_i^\mu \delta^2 (z-z_i)\,.
\end{align}
We can then use the known results from quantum field theory (it here being the $2$--dimensional CFT) to arrive, up to momentum conservation, at
\begin{align}
\begin{aligned}
    \mathcal{E}_3 & \sim \exp \Big(-\tfrac12 \int \md^2 z \md^2 w \, J_\mu(z) \, \langle X^\mu (z) X^\nu(w)\rangle  \, J_\nu(w) \Big) \,.
\end{aligned}
\end{align}
By substituting the current \eqref{eq:current} and the propagator \eqref{eq:diskprop} in this expression, it is straightforward to see that it becomes the result \eqref{eq:kb3} used in the previous Section and it is easy to see that the formula generalizes with the upper bound in the sum corresponding to the number of legs. In fact, $\mathcal{E}_4$, integrated over the vertex operator insertions as previously discussed, namely the $4$--tachyon amplitude, \textit{is} the Veneziano amplitude (the $s$--channel of which is given in) \eqref{eq:ven}!

Computing string amplitudes with massive external states becomes technically hard as the level increases\footnote{The powerful DDF method offers access to highly excited strings, albeit not in a manifestly covariant way.}; the reader is referred to \cite{Feng:2010yx,Lust:2021jps, Lust:2023sfk} for the example of the lightest massive spin-$2$ state. However, a generalization of the current \eqref{eq:current} offers access to the fundamental building block of all tree--level amplitudes with any \textit{subleading trajectories} as external states \cite{Markou:2023ffh}. The generalization is possible via an \textit{exponentiation} of the polynomials of the external legs' vertex operators. For the example of the leading Regge trajectory (where we sum over all its states for convenience),
\begin{align}
V_{\textrm{leading}} = \sum_{s=0}^\infty \frac{1}{s!} \varepsilon_{\mu(s)} \partial X^{\mu_1} \dots \partial X^{\mu_s} e^{\mathrm{i} p\cdot X} \,,\quad p^2 = - \tfrac{s-1}{\alpha'}\,,
\end{align}
 this amounts to rewriting \cite{Kawai:1985xq} (in the notation used here)
\begin{align}
\begin{aligned}
V_{\textrm{leading}} & = \sum_{s=0}^\infty \tfrac{1}{s!}  (\varepsilon \cdot \partial X)^s e^{\mathrm{i} p\cdot X} \\ & = \sum_{s=0}^\infty \tfrac{1}{s!}  \Big(-\mathrm{i} \varepsilon \cdot \frac{\partial}{\partial \xi} \Big)^s e^{\mathrm{i} p\cdot X+ \mathrm{i} \xi \cdot \partial X} \bigg|_{\xi=0} = \exp  \Big(-\mathrm{i} \varepsilon \cdot \frac{\partial}{\partial \xi} \Big) \, e^{\mathrm{i} p\cdot X+ \mathrm{i} \xi \cdot \partial X} \bigg|_{\xi=0}  \,.
\end{aligned}
\end{align}
In the first step of the above, $s$ copies of an auxiliary vector $\varepsilon^\mu$, that satisfies $p \cdot \varepsilon = 0 =\varepsilon \cdot \varepsilon $ (such that $\varepsilon_{\mu(s)}$ be transverse and traceless), have been introduced via $\varepsilon_{\mu(s)} \rightarrow \varepsilon_1 \ldots \varepsilon_s$. In the second step, another auxiliary vector $\xi^\mu$ has been introduced, such that in the end the polynomial is written as a \textit{differential operator} acting on an exponential; after the derivation, the parameter $\xi^\mu$ has to be set to $0$. Let us consider a more complicated trajectory, for example the clone \eqref{eq:clone2} of the leading Regge at $w=2$ and let us focus on its second term. The corresponding term in the vertex operator reads
\begin{align}
\sum_{s=2}^\infty \tfrac{D+2s-1}{s!} \varepsilon_{\mu(s)}  \partial X^{\mu_1} \dots \partial X^{\mu_{s-1}}  \partial^3 X^{\mu_s}e^{\mathrm{i} p\cdot X}  \,,
\end{align}
which we may rewrite as
\begin{align} \label{eq:secondd}
  \sum_{s=2}^\infty \tfrac{D+2s-1}{s!}   \Big(-\mathrm{i} \varepsilon \cdot \frac{\partial}{\partial \xi^{(1)}} \Big)^{s-1}   \Big(-\mathrm{i} \varepsilon \cdot \frac{\partial}{\partial \xi^{(3)}} \Big) e^{\mathrm{i} p\cdot X+ \mathrm{i} \xi^{(1)} \cdot \partial X + \mathrm{i} \xi^{(3)} \cdot \partial^3 X}  \bigg|_{\xi=0}\,,
\end{align}
by introducing \textit{two} auxiliary vectors $\xi^{(1)}$ and $\xi^{(3)}$ which have to be set to $0$ at the end of the derivation. The expression \eqref{eq:secondd} contains \textit{truncated exponents}, that may be represented as an integral of an exponential over a simplex  \cite{Markou:2023ffh}; this a feature of all \textit{truncated clones}.

What have we learned? That to calculate an $N$--point amplitude ($N$ here standing for the number of legs and not the level), we first need the correlator
\begin{align} \label{eq:all}
\langle  \prod_{i=1}^N : \exp\Big(\mathrm{i} p_i \cdot X_i + \mathrm{i} \sum_{n_i=1} \xi_i^{(n_i)} \cdot \partial^{n_i} X_i \Big) :\, \rangle_{\mathbb{D}_2}\,,
\end{align}
where the index $i$ enumerates the external legs and an auxiliary vector $\xi_\mu^{(n_i)}$ is introduced for every descendant $\partial^{n_i}X_i^\mu$ that may be excited in the vertex operator of the $i$--th leg. Using the generating functional method, the correlator \eqref{eq:all} can be computed explicitly and the result is given in \cite{Markou:2023ffh}. It then suffices to act on this result with the differential operators of the external legs (and perform the worldsheet integration for more than $3$ legs) to compute the amplitude! Several examples are given in \cite{Markou:2023ffh}. Let us finally note that for trajectories with more than one row, a different auxiliary vector $\varepsilon_j^{\mu}$ has to be introduced for every row $j$, with $\varepsilon_i \cdot \varepsilon_j =0 = p \cdot \varepsilon_j$, while the Young symmetry must be imposed at the end of the calculation.

%%%%%%%%%
\section{Conclusions}
\label{sec:concl}
%%%%%%%%%

In these lectures, we have reviewed a selection of elements of classical and quantum bosonic string propagation, the level--by--level construction of the physical string spectrum, CFT and string amplitudes basics,  as well as the new, covariant and efficient method of \cite{Markou:2023ffh} for the algorithmic construction of entire subleading Regge trajectories and the calculation of their tree--level amplitudes. The non--trivial extension to the superstring was developed in \cite{Basile:2024uxn} and is beyond the scope of these lectures. The main message that the author hopes to have shared is that our understanding of the deep string spectrum is \textit{tied} to our understanding of the origin of the good UV behavior of string scattering amplitudes, namely a feature of string theory that  has kept it a promising framework towards a quantum theory of gravity through the decades. Consequently, employing the new technology to decode underlying aspects of string amplitudes is a fascinating direction, that could perhaps be related to questions of \textit{uniqueness} of string theory addressed for example in \cite{Cheung:2022mkw}. Another interesting direction would be the further development of the technology and the combination of its use with non--perturbative objects (branes), with a vision of contributing to our understanding of the black hole information paradox in the context of string theory. And let us leave the reader with a question: can the technology serve towards realizing the infinitely many massive higher--spin string states as originating,  via a mechanism of spontaneous symmetry breaking, in a higher--spin gravity of infinitely many massless higher--spins, as long speculated within the high energy community?

\paragraph{Acknowledgements.} The author warmly thanks the organizers of the 2024 Modave Summer School in Mathematical Physics, especially Noémie Parrini, for the kind invitation to deliver these lectures, as well as the participating students for raising sharp questions. The author is very grateful to Evgeny Skvortsov, from whom she learned a great deal, for enlightening discussions and a formative collaboration in the work reviewed here; she further thanks Thomas Basile, Paolo Di Vecchia and Augusto Sagnotti for illuminating discussions, as well as the Institut des Hautes \'Etudes Scientifiques for warm hospitality at the proofreading stage of these notes. The author is supported by a fellowship of the Scuola Normale Superiore and by INFN (I.S. GSS-Pi).

\end{document}